\documentclass[journal,comsoc]{IEEEtran}

\usepackage{framed}
\usepackage{hyperref}
\usepackage{tcolorbox}
\usepackage{amsmath}
\usepackage{mathtools}
\usepackage{bbm}
\usepackage[ruled,linesnumbered]{algorithm2e}
\usepackage{bm}
\usepackage{stackrel}
\usepackage{subfig}
\usepackage{epsf}
\usepackage{amsmath,amssymb}

\usepackage{graphicx}
\usepackage{color}
\usepackage{cite}
\usepackage{multirow,tabularx}
\usepackage{ifthen}
\usepackage{epstopdf}
\input{epstopdf.sty}

\usepackage{times}
\input{epsf.sty}

\makeatletter
\def\BState{\State\hskip-\ALG@thistlm}
\makeatother

\newcommand{\hide}[1]{\ifthenelse{\boolean{false}}{#1}{}}

\newtheorem{theorem}{{\bf Theorem}}

\newtheorem{lemma}{{\bf Lemma}}
\newtheorem{assumption}{{\bf Assumption}}

\newcommand{\qed}{\nobreak \ifvmode \relax \else
      \ifdim\lastskip<1.5em \hskip-\lastskip
      \hskip1.5em plus0em minus0.5em \fi \nobreak
      \vrule height0.75em width0.5em depth0.25em\fi}

\newcommand{\beq}{\begin{equation}}
\newcommand{\eeq}{\end{equation}}
\newcommand{\barr}{\begin{array}}
\newcommand{\earr}{\end{array}}

\newcommand{\benum}{\begin{enumerate}}
\newcommand{\eenum}{\end{enumerate}}

\newcommand{\bit}{\begin{itemize}}
\newcommand{\eit}{\end{itemize}}

\newcommand{\bc}{\begin{center}}
\newcommand{\ec}{\end{center}}

\newcommand{\bdes}{\begin{description}}
\newcommand{\edes}{\end{description}}

\newcommand{\bfig}{\begin{figure}}
\newcommand{\efig}{\end{figure}}

\newcommand{\bemq}{\begin{quote} \begin{em}}
\newcommand{\eemq}{\end{em} \end{quote}}

\newcommand{\bmp}{\begin{minipage}}
\newcommand{\emp}{\end{minipage}}

\newcommand{\bsp}{\begin{slide*}}
\newcommand{\esp}{\end{slide*}}
\newcommand{\bsl}{\begin{slide}}
\newcommand{\esl}{\end{slide}}

\newcommand{\blem}{\begin{lemma}}
\newcommand{\elem}{\end{lemma}}
\newcommand{\bthm}{\begin{theorem}}
\newcommand{\ethm}{\end{theorem}}

\IEEEoverridecommandlockouts
\SetKwFor{For}{At every source}{do}{~}

\begin{document}

\title{Fresh-CSMA: A Distributed Protocol for Minimizing Age of Information}
\author{Vishrant Tripathi, Nicholas Jones, and Eytan Modiano
\IEEEcompsocitemizethanks{\IEEEcompsocthanksitem Vishrant Tripathi, Nicholas Jones, and Eytan Modiano are with the Laboratory for Information and Decision Systems (LIDS), Massachusetts Institute of Technology, Cambridge, MA, 02139. A preliminary version of this paper appeared in the conference proceedings of IEEE INFOCOM 2023.\protect\\
E-mail: \{vishrant, jonesn, modiano\}@mit.edu.
}
}

\IEEEaftertitletext{\vspace{-0.6\baselineskip}}

\maketitle
\begin{abstract}
We consider the design of distributed scheduling algorithms that minimize Age of Information (AoI) in single-hop wireless networks. The centralized max-weight policy is known to be nearly optimal in this setting; hence, our goal is to design a distributed Carrier Sense Multiple Access (CSMA) scheme that can mimic its performance. To that end, we propose a distributed protocol called Fresh-CSMA and show that in an idealized setting, Fresh-CSMA can match the scheduling decisions of the max-weight policy with high probability in each frame, and also match the theoretical performance guarantees of the max-weight policy over the entire time horizon. We then consider a more realistic setting and study the impact of protocol parameters on the probability of collisions and the overhead caused by the distributed nature of the protocol. We also consider the monitoring of Markov sources and extend our approach to CSMA protocols that incorporate Age of Incorrect Information (AoII) instead of AoI. Finally, we provide simulations that support our theoretical results and show that the performance gap between the ideal and realistic versions of Fresh-CSMA is small.
\end{abstract}

\section{Introduction}
\label{sec:intro}

Many emerging applications require timely delivery of information updates over communication networks. 
Age of Information (AoI) is a metric that captures this notion of timeliness of received information at a destination \cite{kaul2012real,kam2013age,yin17_tit_update_or_wait}. Unlike packet delay, AoI measures the lag in obtaining information at a destination node, and is therefore suited for applications involving time sensitive updates. Age of information, at a destination, is defined as the time that has elapsed since the last received information update was generated at the source. AoI, upon reception of a new update, drops to the time that has elapsed since generation of the update, and grows linearly otherwise. 


The design of scheduling policies to minimize age of information metrics over single-hop wireless networks has received special interest over the past few years. In \cite{kadota2018scheduling, kadota2018scheduling2}, the authors consider a single-hop broadcast setting with the aim of minimizing weighted sum AoI. In \cite{talak2018optimizing}, the authors also study the minimization of the weighted sum of AoI but under general interference constraints. These lines of work prove constant factor optimality of three different classes of scheduling policies; stationary randomized, max-weight and Whittle index. In \cite{maatouk2020optimality}, the authors prove asymptotic optimality of the Whittle index policy and in \cite{li2019general}, the authors extend prior results to different sampling behaviors, update sizes and transmission times in a single-hop broadcast setting. In \cite{jhun2018age, tripathi2019whittle}, the authors extend the AoI minimization framework to consider general nonlinear cost functions of AoI. 

Importantly, all of the works above focus on the design of \textit{centralized scheduling algorithms}. Specifically, at the beginning of each time-slot, the base station looks at the AoI values for each node in the network and then decides which node to poll for an update. This requires support for polling protocols, which might not be available at the MAC layer and might involve excessive overhead for networks with many nodes. This has motivated the need to study distributed schemes for information freshness in wireless networks.

In \cite{talak2018distributed}, the authors consider a simple class of distributed algorithms - each node transmits using a fixed attempt probability in each time-slot, and they design a scheme to find the attempt probabilities that minimize weighted sum AoI. In \cite{jiang2018can}, the authors consider a single-hop setting with stochastic arrivals and solve the AoI minimization problem by deriving a Whittle index and propose a heuristic ALOHA-like scheme called Index-Prioritized-Random-Access (IPRA) where a node is active with a fixed probability but only when its AoI exceeds a specified threshold. The idea of ALOHA with thresholds has been explored in further detail in subsequent works. In \cite{chen2022age, chen2021real}, the authors study the performance of ALOHA style random access protocols for information freshness and propose the idea of ``thinning", where only nodes with AoI greater than a specified threshold remain active. Along similar lines, the performance of threshold-ALOHA for AoI minimization is also analyzed in \cite{yavascan2020analysis,yavascan2021analysis} with performance bounds derived in a symmetric setting. 

Another class of distributed protocols commonly used in wireless networks for medium access is Carrier Sense Multiple Access with Collision Avoidance, also known as CSMA/CA. Throughout this work, when we use the term CSMA, we use it to denote CSMA/CA style protocols. Age of information has also been analyzed in settings where nodes employ CSMA \cite{maatouk2020CSMA, baiocchi2020age, wang2019broadcast, chen2022age, bedewy2020optimizing, bedewy2021low}. In \cite{maatouk2020CSMA}, the authors analyze an idealized version of IEEE 802.11 CSMA and optimize the backoff timer parameters to minimize AoI. They show that this version of CSMA has poor delay and freshness performance in certain settings and suggest the need for new distributed scheduling schemes for AoI. In \cite{wang2019broadcast, baiocchi2020age}, the authors analyze AoI under standard CSMA in broadcast environments. In \cite{bedewy2020optimizing,bedewy2021low}, the authors study sleep-wake carrier sensing based scheduling with the goal of minimizing energy consumption together with AoI.

The CSMA protocol has been well studied in wireless networks for a long time, especially for optimizing throughput and utility. It was shown in \cite{kleinrock1975packet} that CSMA tends to outperform ALOHA in terms of both throughput and delay. In \cite{bianchi2000performance}, the author developed an approximation that allows closed-form analysis for the IEEE 802.11 implementation of CSMA. More recently, there has been work on throughput and utility optimization by trying to replicate centralized scheduling policies' behavior using CSMA style schemes \cite{jiang2009distributed, jiang2012fast, ni2011q, li2013optimal}. Typically, these works involve modifying the way CSMA backoff timers work by adding dependence on the current network state (e.g. queue lengths) and then analyzing performance guarantees by comparing to a centralized scheduling scheme. Our analysis and approach are motivated by this line of work, in particular the Fast-CSMA protocol proposed in \cite{li2013optimal}.

In this work, we propose Fresh-CSMA to replicate the behavior of centralized scheduling schemes that minimize AoI. In Section \ref{sec:model} we discuss our system model and set up the single-hop weighted age minimization problem. In Section \ref{sec:ideal} we introduce the Fresh-CSMA protocol in an idealized setting and provide performance guarantees that show that it can closely match the centralized max-weight scheduling policy both per time-slot and over the entire time horizon. In Section \ref{sec:semi_real}, we relax some of the assumptions from our idealized model and study the Fresh-CSMA protocol under a more realistic setting. We analyze two keys aspects -  the probability of collision and the total time lost due to the backoff timers during which the channel remains idle. In Section \ref{sec:AoII}, we consider the recently proposed information freshness metric called Age of Incorrect Information (AoII) and extend our CSMA design to incorporate this metric. In Section \ref{sec:sim}, we provide simulations that support our theoretical results.

A preliminary version of our paper appeared in the conference proceedings of IEEE INFOCOM 2023.


\section{System Model}
\label{sec:model}
\begin{figure}
	\centering
	\includegraphics[width=0.8\linewidth]{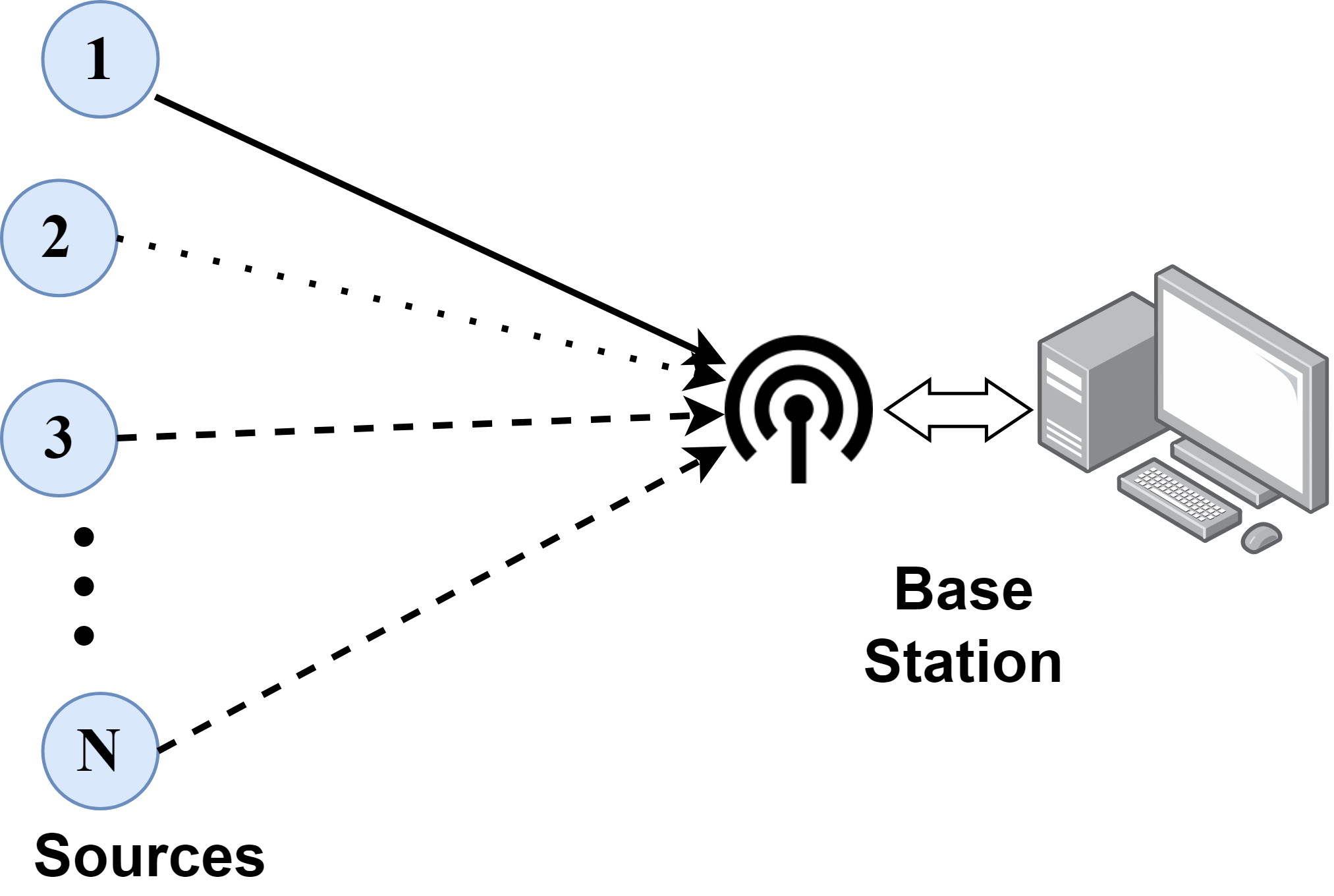}
	\caption{Single-hop broadcast network with $N$ sources sending updates to a base station over a shared channel.}
	\label{fig:model}
\end{figure}
Consider a single-hop wireless network with $N$ sources generating real-time status updates that need to be sent to a monitoring base station (see Fig.~\ref{fig:model}). We consider a slotted system in which each source takes a single time-slot to transmit an update to the base station. Due to interference, only one of the sources can transmit successfully in any given time-slot. If multiple sources decide to transmit, a collision occurs and the transmitted updates are lost. 

For every source $i$, the age of information at the base station $A_i(t)$ measures the time elapsed since it received a fresh information update from the source. We assume active sources, i.e. in any time-slot, sources can generate fresh updates at will.  Let $s(t)$ be the set of sources transmitting in time-slot $t$. Then, the age of information $A_i(t)$ evolves as:
\begin{equation}
\label{AoI_def}
A_i(t+1) =
 \begin{cases}
      1, &  \text{if }s(t) = \{i\},\\
      A_i(t)+1, & \text{otherwise.}
 \end{cases}
\end{equation}

The metric of interest in this work will be average AoI, which is simply the long-term time-average of the AoI process. Specifically,
\begin{equation}
\label{eq:avg_AoI_def}
    \bar{A}_i \triangleq \limsup_{T \rightarrow \infty} \frac{1}{T}  \sum_{t=1}^{T} A_i(t) .
\end{equation}

The goal of this work is to design a \textit{distributed} wireless scheduling policy that minimizes the weighted sum of average AoI across all sources:
\begin{equation}
\label{eq:age_opt_prob}
\underset{\pi}{\operatorname{argmin}} \bigg( \limsup_{T \rightarrow \infty}   \bigg[\frac{1}{T} \sum_{t = 1}^{T} \sum_{i=1}^{N} w_i A_{i}(t) \bigg] \bigg).
\end{equation}
Here, the weights $\{w_1, w_2, ..., w_N\}$ are positive \textit{integers} that denote the relative importance of each source to the overall monitoring or control application.

\section{Distributed Scheduling Design}
\label{sec:ideal}
Before we introduce our distributed scheduling design, we briefly discuss key results from prior works that have looked at the same problem but from a centralized perspective. 


In \cite{kadota2018scheduling}, the authors considered a similar single-hop network setting with the goal of minimizing weighted-sum average AoI, i.e. solving \eqref{eq:age_opt_prob}. First, they considered the class of stationary randomized policies. Each policy within this class is simply a probability distribution over the set of sources and the scheduling decision is sampled from this distribution i.i.d. at the beginning of every time-slot. They showed that under the optimal stationary randomized policies that solve \eqref{eq:age_opt_prob}, each source $i$ is scheduled with probability
\begin{equation}
\label{eq:opt_sr}
  \pi^{*}_i = \frac{\sqrt{w_i}}{\sum_{i=1}^{N}\sqrt{w_i}}, \forall i \in [N].  
\end{equation}
They further showed that the best stationary randomized policies can be at most a factor of two away from the overall optimal policy.

They also proposed a centralized policy motivated by Lyapunov drift arguments called the max-weight policy. This policy, under reliable channels, makes scheduling decisions $\pi^{mw}(t)$ as follows:
\begin{equation}
\label{eq:max_weight}
    \pi^{mw}(t) = \underset{j \in [N]}{\operatorname{argmax}}\bigg(  w_j A^{2}_j(t)\bigg).
\end{equation}
In \cite{AoI_book}, the authors showed that this policy is at most a factor of two away from optimal using Lyapunov drift arguments. However, unlike stationary randomized policies, this policy turns out to be \textit{nearly optimal} in practice.

The goal of our distributed design is to replicate the decision making and performance guarantees of max-weight policies of the form \eqref{eq:max_weight}.

\subsection{Fresh-CSMA}
In general, a CSMA/CA style protocol involves the following steps. First, a node senses the channel to see if it is free. If the node determines the channel to be available, it starts transmitting a packet. Otherwise, if the channel is occupied, it generates a random backoff time and starts a timer counting down from this value. During this period, the node continuously senses the channel and only counts down \textit{when the channel is free}. Once the timer hits zero, the node transmits a frame. Note that the timer can only hit zero when the channel is known to be free. Depending on whether an ACK is received or not from the receiving station, the node updates its random backoff timer parameters for the next transmission.

The details of how to sample the backoff times, how to update them in case of a collision or re-transmission, and how to implement channel sensing determine the exact flavor of CSMA being implemented. To develop our scheme and provide tractable analysis, we will consider an idealized channel sensing setup that is commonly used in theoretical works addressing CSMA \cite{jiang2009distributed, jiang2012fast, ni2011q, li2013optimal}. This involves making the following key assumptions.
\begin{framed}
\begin{assumption}
\label{ass:timers}
Backoff timers are implemented in continuous time.
\end{assumption}
\begin{assumption}
\label{ass:sensing}
Carrier sensing happens instantly.
\end{assumption}
\begin{assumption}
\label{ass:slotted}
There is a discrete slotted system and all nodes start their backoff timers at the beginning of each time-slot.
\end{assumption}
\begin{assumption}
\label{ass:overhead}
Backoff timers are implemented with arbitrary precision, and can be made negligible in comparison to the duration of a time-slot.
\end{assumption}
\end{framed}


Under these assumptions, a version of the classic CSMA protocol that uses exponential backoff timers is described in Alg.~\ref{alg:i-CSMA}. We use $t$ to denote the discrete time-slots and $\tau$ to denote continuous time within each time-slot. We normalize the time-slot length to be $1$ so $\tau$ is a continuous timer that increases from $0$ to $1$ within each time-slot. 
\begin{algorithm}
	\DontPrintSemicolon
	%
	\SetKwInOut{Input}{Input}\SetKwInOut{Output}{Output}
	\Input{parameter $\alpha > 1$}
	\While{ $t \in 1,...,T$ }{
		\For{$i \in 1,...,N$}{
		Generate a  random timer $Z_i(t) \sim \text{exp}(\alpha)$.\; 
		\While{$\tau < Z_i(t)$}{
		Stay silent\;
		}
		\If{Channel is free}{
		Transmit\;
		}
		}
	}		
	\caption{Idealized CSMA}
	\label{alg:i-CSMA}
\end{algorithm}

As a consequence of our idealized assumptions, note that a packet collision happens only if two nodes choose the exact same backoff times. Since the probability that two exponential random variables take the exact same value is zero, so the probability of packet collisions is also zero in this idealized setup. 

The protocol above consists of two key ideas. First, each source $i$ generates a  random timer $Z_i(t)$ that is i.i.d. exponentially distributed. Second, the source with the timer that runs out first gets to transmit in the entire time-slot $t$, i.e. 
\begin{equation}
    \label{eq:min_timer}
    \pi(t) = \underset{i \in [N]}{\operatorname{argmin}}~~  Z_i(t).
\end{equation}
Due to Assumption \ref{ass:overhead}, we can scale the backoff timers by a factor $\delta$ such that they are negligible in comparison to the slot length, i.e. $\delta Z_i(t) \ll 1, \forall i$ with high probability. 

Next, we modify this CSMA protocol to create Fresh-CSMA, described in Alg.~\ref{alg:i-Fresh-CSMA}. This is also an idealized distributed protocol, but with the goal of replicating the behavior of the max-weight scheduling policy \eqref{eq:max_weight} for AoI minimization. This style of CSMA is motivated by the fast-CSMA protocol proposed in \cite{li2013optimal}. 
\begin{algorithm}
	\DontPrintSemicolon
	\SetKwInOut{Input}{Input}\SetKwInOut{Output}{Output}
	\Input{parameters $\alpha > 1, \delta \ll 1$}
	\While{ $t \in 1,...,T$ }{
		\For{$i \in 1,...,N$}{
		Generate a  random timer $Z_i(t) \sim \text{exp}\bigg(\alpha^{w_i A^2_i(t)}\bigg)$.\; 
		\While{$\tau < \delta Z_i(t)$}{
		Stay silent\;
		}
		\If{Channel is free}{
		Transmit\;
		}
		}
	}
	\caption{Idealized Fresh-CSMA}
	\label{alg:i-Fresh-CSMA}
\end{algorithm}

Note that Fresh-CSMA consists of two key steps. First, each source $i$ generates a  random timer $\delta Z_i(t)$ where $Z_i(t)$ is exponentially distributed with the parameter $\alpha^{w_i A^2_i(t)}$. Then, the source with the timer that runs out first gets to transmit in the entire time-slot $t$, i.e. according to \eqref{eq:min_timer}. 

Importantly, each source only requires knowledge of its own AoI and scheduling weight to compute the backoff timer, thus maintaining the distributed nature of the protocol. The following lemma describes the structure of scheduling decisions made by this scheduling scheme. 
\begin{framed}
\begin{lemma}
For Fresh-CSMA at time-slot $t$, the probability that source $i$ is scheduled is given by:
\begin{equation}
    \label{eq:aoi_csma_probability}
    r_i(t) = \frac{\alpha^{w_i A^2_i(t)}}{\sum\limits_{j=1}^{N}\alpha^{w_j A^2_j(t)} }.
\end{equation}
\end{lemma}
\end{framed}
\begin{IEEEproof}
First, note that the scheduling decision at time-slot $t$ is given by \eqref{eq:min_timer} where $Z_i(t)$ are independent and exponentially distributed with the parameters $\alpha^{w_i  A^2_i(t)}$. Let $Z(t) \triangleq \underset{i \in [N]}{\operatorname{min}}~~  Z_i(t)$. Then, $Z(t)$ is the minimum of $N$ independent exponential random variables. Thus, it is also exponentially distributed and with the parameter $\sum_{j=1}^{N}\alpha^{w_j A^2_j(t)}$. Let $\lambda_i \triangleq \alpha^{w_i A^2_i(t)}$. We are interested in calculating the probability 
\begin{align}
\begin{aligned}
     r_i(t) &= \mathbb{P}\big(\pi(t) = i\big) = \mathbb{P}\big(Z(t) = Z_i(t)\big) \\
     &= \int_{0}^{\infty} f_{Z_i(t)}(x) \prod_{k \neq i} \mathbb{P}\big(Z_k(t) > x\big) dx \\
     &= \int_{0}^{\infty} \lambda_i e^{-\lambda_i x}  \prod_{k \neq i} e^{-\lambda_k x} dx\\
     &= \frac{\lambda_i}{\sum_{k=1}^{N} \lambda_k}.
\end{aligned}
\end{align}
This completes the proof.
\end{IEEEproof}

Using Lemma 1, we next show that if the parameter $\alpha$ is set to be large enough, then in any particular time-slot, Fresh-CSMA will make the same scheduling decision as the max-weight policy with high probability.
\begin{framed}
\begin{theorem}
\label{thm:csma_timeslot}
Given any $\delta \in (0,1)$ and $A_1(t),...,A_N(t)$; if we set $\alpha \geq (N-1)\frac{1-\delta}{\delta}$, then the following holds
\begin{equation}
    \mathbb{P}\bigg(\pi^{Fresh-CSMA}(t) = \pi^{mw}(t)\bigg) \geq 1-\delta.
\end{equation}
Here, $\pi^{mw}(t)$ is the max-weight scheduling decision given by \eqref{eq:max_weight}, while $\pi^{Fresh-CSMA}(t)$ is the scheduling decision made by Fresh-CSMA. 
\end{theorem}
\end{framed}
\begin{IEEEproof}
We divide the proof into two parts.

\textbf{Case 1:} The expression $\underset{j \in [N]}{\operatorname{argmax}}\bigg(  w_j A^{2}_j(t)\bigg)$ has a unique maximum. Let this maximum be the source $1$ without loss of generality. Then, the max-weight decision is to schedule source $1$. Since we have assumed all the weights $w_i$ to be positive integers and the AoIs $A_i(t)$ are integers by definition, so the quantities $w_i A^{2}_i(t)$ are also positive integers and the following must hold:
\begin{equation}
    \label{eq:th1_c1_ineq}
    w_1 A^2_1(t) - w_i A^2_i(t) \geq 1, \forall i \neq 1.
\end{equation}
Note that in the special case of $w_i = 1, \forall i$, \eqref{eq:th1_c1_ineq} holds when the AoIs across the different sources are unique.

Now, applying Lemma 1, we can calculate the following probability
\begin{align}
\begin{aligned}
    \mathbb{P}\bigg(\pi^{Fresh-CSMA}(t) = 1\bigg) &= \frac{\alpha^{w_1 A^2_1(t)}}{\sum\limits_{j=1}^{N}\alpha^{w_j A^2_j(t)} }\\
        &= \frac{1}{ 1 + \sum_{i=2}^{N} \alpha^{w_i A^2_i(t) - w_1 A^2_1(t)}}\\
        &\geq \frac{1}{1+(N-1)\alpha^{-1}}\\
        &\geq 1-\delta.
\end{aligned}
\end{align}
The first inequality follows by using \eqref{eq:th1_c1_ineq} while the second inequality follows due to the fact that $\alpha \geq (N-1)\frac{1-\delta}{\delta}$.

\textbf{Case 2:} The expression $\underset{j \in [N]}{\operatorname{argmax}}\bigg(  w_j A^{2}_j(t)\bigg)$ has multiple maxima. 
Suppose that the set of maxima is given by the nodes $\{1,...,k\}$. Then, the max-weight policy will choose one of these sources to be scheduled. We want to lower bound the probability that Fresh-CSMA chooses a node from within this set. To do so, we first make a similar observation as in the case above.
\begin{equation}
    \label{eq:th1_c2_ineq}
    w_1 A^2_1(t) - w_j A^2_j(t) \geq 1,  \forall j = k+1,...,N.
\end{equation}
Using Lemma 1, we calculate the probability of interest
\begin{align}
\begin{aligned}
    \mathbb{P}\bigg( \pi^{Fresh-CSMA}(t) &\in \{1,...,k \} \bigg) = \frac{k \alpha^{w_1 A^2_1(t)}}{\sum\limits_{j=1}^{N}\alpha^{w_j A^2_j(t)} } \\
    &= \frac{k}{ k + \sum_{i=k+1}^{N} \alpha^{w_i A^2_i(t) - w_1 A^2_1(t)}}\\
        &\geq \frac{1}{1+(N-k)\alpha^{-1}}\\
        &\geq 1-\delta.
\end{aligned}
\end{align}
As before, the first inequality follows by using \eqref{eq:th1_c2_ineq} while the second inequality follows due to the fact that $\alpha \geq (N-1)\frac{1-\delta}{\delta}$. This completes the proof.
\end{IEEEproof}

Next, we show that our idealized Fresh-CSMA protocol has the same theoretical long-term performance guarantees as the max-weight policy.
\begin{framed}
\begin{theorem}
\label{thm:csma_opt}
Given any set of integer weights $w_1,...,w_N$, if we set $\alpha > \bigg( \frac{(N-1)\sum_{i=1}^{N} \sqrt{w_i}}{\min_{j}\sqrt{w_j}} \bigg)$, then the following holds:
\begin{equation}
    \frac{ \sum_{i=1}^{N} w_i \bar{A}^{csma}_i }{ \sum_{i=1}^{N} w_i \bar{A}^{opt}_i } \leq 2.
\end{equation}
Here, $\bar{A}^{csma}_i$ is the average AoI for source $i$ under the Fresh-CSMA policy while $\bar{A}^{opt}_i$ is the average AoI of source $i$ under an optimal policy $\pi^{opt}$ that solves the age minimization problem \eqref{eq:age_opt_prob}.
\end{theorem}
\end{framed}

\begin{IEEEproof}
Consider the linear Lyapunov function as defined below:
\begin{equation}
    L(t) \triangleq \sum_{i=1}^{N} \sqrt{w_i} A_i(t).
\end{equation}
We can define the one-slot Lyapunov drift 
$
    \Delta(t) \triangleq L(t+1) - L(t).
$
The main challenge in proving the performance bound above, is to first show an intermediate result relating the Lyapunov drift of the Fresh-CSMA policy to that of the optimal stationary randomized policy described by \eqref{eq:opt_sr}. 
\begin{lemma}
\label{lem:drift}
Consider any $\bm{A(t)} = \{A_1,...,A_N(t)\}$. Let $\Delta^{csma}(t)$ be the one-slot Lyapunov drift of the Fresh-CSMA policy and $\Delta^{sr}(t)$ be the one-slot Lyapunov drift of the optimal stationary randomized policy. Then, the following holds: 
\begin{equation}
\label{eq:drift_comp}
    \mathbb{E}\bigg[\Delta^{csma}(t) \bigg| \bm{A(t)}\bigg] \leq \mathbb{E}\bigg[\Delta^{sr}(t) \bigg| \bm{A(t)}\bigg], \forall \bm{A(t)}.
\end{equation}
\end{lemma}
 \begin{IEEEproof}
 We first calculate an expression for the drift of the Fresh-CSMA policy. Recall that $r_j(t) \triangleq \mathbb{P}\big(\pi^{AoI-CMSA}(t) = j\big)$ and is given by \eqref{eq:aoi_csma_probability}.
\label{pf:lem_drift}
\begin{align}
\begin{aligned}
\label{eq:drift_csma}
    \mathbb{E}&\bigg[\Delta^{csma}(t) \bigg| \bm{A(t)}\bigg] \\ =&\sum_{j=1}^{N} r_j(t) \bigg(\sqrt{w_j} + \sum_{i \neq j}\sqrt{w_i} (A_i(t)+1) \bigg) - \sum_{j=1}^{N} \sqrt{w_j} A_j(t)\\
    =&\sum_{j=1}^{N} \sqrt{w_j} - \sum_{j=1}^{N} r_j(t)\sqrt{w_j}A_j(t).
\end{aligned}
\end{align}

Repeating the above steps for the optimal stationary randomized policy, we get: 
\begin{align}
\begin{aligned}
\label{eq:drift_sr}
    \mathbb{E}\bigg[\Delta^{sr}(t) \bigg| \bm{A(t)}\bigg] 
    = \sum_{j=1}^{N} \sqrt{w_j} - \sum_{j=1}^{N} \pi^{*}_j\sqrt{w_j}A_j(t).
\end{aligned}
\end{align}
Recall that $\pi^*_j$ are scheduling probabilities for the optimal stationary randomized policy, given by \eqref{eq:opt_sr}.

Consider the difference between \eqref{eq:drift_csma} and \eqref{eq:drift_sr}

\begin{align}
    \begin{aligned}
    \label{eq:drift_diff_1}
    \mathbb{E}\bigg[&\Delta^{csma}(t) \bigg| \bm{A(t)}\bigg] - \mathbb{E}\bigg[\Delta^{sr}(t) \bigg| \bm{A(t)}\bigg] \\
    &= \sum_{j=1}^{N} \sqrt{w_j} A_j(t) \big(\pi^{*}_j - r_j(t) \big)\\
    &= \sum_{j=1}^{N} \sqrt{w_j} A_j(t) \bigg( \frac{\sqrt{w_j}}{\sum\limits_{i=1}^{N}\sqrt{w_i}} - \frac{\alpha^{w_j A^2_j(t)}}{\sum\limits_{i=1}^{N}\alpha^{w_i A^2_i(t)} } \bigg)
    \end{aligned}
\end{align}

Note that we are only interested in the sign of \eqref{eq:drift_diff_1}, so we can instead look at
\begin{align}
\begin{aligned}
    &\sum\limits_{j=1}^{N} \sqrt{w_j} A_j(t) \bigg( \sqrt{w_j}  (\sum\limits_{i=1}^{N}\alpha^{w_i A^2_i(t)}) - \alpha^{w_j A^2_j(t)} (\sum\limits_{i=1}^{N}\sqrt{w_i})     \bigg)\\
    &= \sum_{j=1}^{N} \alpha^{w_j A^2_j(t)} \bigg( \sum_{i=1}^{N} w_i A_i(t) -  \sqrt{w_j} A_j(t) \sum_{i=1}^{N} \sqrt{w_i}   \bigg)\\
    &= \sum_{j=1}^{N} \alpha^{w_j A^2_j(t)} \bigg( \sum_{i=1}^{N} \sqrt{w_i} \big(\sqrt{w_i} A_i(t) -  \sqrt{w_j} A_j(t)\big)   \bigg)
\end{aligned}
\end{align}

Consider without loss of generality that sources are numbered such that $\sqrt{w_1}A_1(t) \geq \sqrt{w_2}A_2(t) \geq ... \geq \sqrt{w_N}A_N(t)$. This automatically implies that source $1$ has the largest value of $\sqrt{w_j}A_j(t)$ among all sources. 

\textbf{Case 1:} First we consider the case when Source $1$ is the unique max-weight scheduling decision, i.e.
$\pi^{mw}(t) = \underset{j \in [N]}{\operatorname{argmax}}\bigg(  {w_j} A^2_j(t)\bigg) = 1.$ Since we have assumed weights $w_i \in \mathbb{Z}^{+}$ and AoIs $A_i(t)$ are also positive integers, the above equation implies that
\begin{equation}
\label{eq:unique}
    {w_1}A^2_1(t) - {w_i} A^2_i(t) \geq 1, \forall i \neq 1.
\end{equation}
This is because $w_i A^2_i(t) \in \mathbb{Z}^{+}$ for all sources $i$. Further, note that $f(x) = \sqrt{x}$ is Lipschitz for $x \in [1,\infty)$ with the Lipschitz constant $0.5$. Applying this fact to \eqref{eq:unique}, we get
\begin{equation}
    \label{eq:unique_sqrt}
    \sqrt{w_1}A_1(t) - \sqrt{w_i} A_i(t) \geq 0.5, \forall i \neq 1.
\end{equation}

Next, we define the following quantities
\begin{equation}
    \gamma_j \triangleq \bigg( \sum_{i=1}^{N} \sqrt{w_i} \big(\sqrt{w_i} A_i(t) -  \sqrt{w_j} A_j(t)\big)   \bigg).
\end{equation}
Using \eqref{eq:unique_sqrt}, it is easy to see that $\gamma_1 \leq -0.5 \sum_{i=1}^{N} \sqrt{w_i} $. We will bound the rest of the values $\gamma_i$ in comparison to $\gamma_1$. 
\begin{equation}
    \begin{split}
    \gamma_j &= \bigg( \sum_{i=1}^{N} \sqrt{w_i} \big(\sqrt{w_i} A_i(t) -  \sqrt{w_j} A_j(t)\big)   \bigg) \\
    &\leq \bigg( \sqrt{w_1} (\sqrt{w_1} A_1(t) -  \sqrt{w_j} A_j(t)) \\&+ \sum_{i=2}^{j-1} \sqrt{w_i} \big(\sqrt{w_1} A_1(t) -  \sqrt{w_j} A_j(t)\big)\\
    &+ \sum_{i=j}^{N} \sqrt{w_i} \big(\sqrt{w_i} A_i(t) -  \sqrt{w_j} A_j(t)\big)
    \bigg)\\
    &\leq \frac{\sum_{i=1}^{N} \sqrt{w_i}}{\sqrt{w_j}} \bigg( \sqrt{w_j}  \big( \sqrt{w_1}A_1(t) - \sqrt{w_j}A_j(t) \big) \\ &+ \sum_{i \neq j} \sqrt{w_i}  \big( \sqrt{w_1}A_1(t) - \sqrt{w_i}A_i(t) \big) \bigg)\\
    &\leq \frac{\sum_{i=1}^{N} \sqrt{w_i}}{\min_{j}\sqrt{w_j}} \big| \gamma_1 \big|, \forall j\neq 1.
    \end{split}
\end{equation}


Using this result, \eqref{eq:unique} and the definition of $\gamma_i$ we get
\begin{align}
    \begin{aligned}
    \label{eq:exp_sum}
    &\sum\limits_{j=1}^{N} \alpha^{w_j A^2_j(t)} \bigg( \sum_{i=1}^{N} \sqrt{w_i} \big(\sqrt{w_i} A_i(t) -  \sqrt{w_j} A_j(t)\big)   \bigg) \\
    &\leq \alpha^{w_1 A^2_1(t)} \big( \gamma_1 \big)  + \sum_{i=2}^{N} \alpha^{w_2 A^2_2(t)} \frac{\sum_{i=1}^{N} \sqrt{w_i}}{\min_{j}\sqrt{w_j}} \big| \gamma_1 \big| \\
    &\leq \alpha^{w_1 A^2_1(t)} \big| \gamma_1 \big| \bigg( -1 + \alpha^{-1}\frac{(N-1)\sum_{i=1}^{N} \sqrt{w_i}}{\min_{j}\sqrt{w_j}} \bigg).  
    \end{aligned}
\end{align}
Clearly, choosing 
\begin{equation}
    \alpha > \bigg( \frac{(N-1)\sum_{i=1}^{N} \sqrt{w_i}}{\min_{j}\sqrt{w_j}} \bigg)
\end{equation}
is sufficient to guarantee that \eqref{eq:exp_sum} is negative and hence \eqref{eq:drift_diff_1} is negative. 
Thus, for this choice of $\alpha$, we observe that
\begin{equation}
    \mathbb{E}\bigg[\Delta^{CSMA}(t) \bigg| \bm{A(t)}\bigg] \leq \mathbb{E}\bigg[\Delta^{SR}(t) \bigg| \bm{A(t)}\bigg], \forall \bm{A(t)}.
\end{equation}
\textbf{Case 2:} Sources $1,...,k$ are all solutions to the following maximization
\begin{equation}
    \pi^{mw}(t) = \underset{j \in [N]}{\operatorname{argmax}}\bigg(  {w_j} A^2_j(t)\bigg) \in \{1,...,k\}.
\end{equation}

We can repeat the exact same analysis as Case 1, but starting with 
\begin{equation}
\label{eq:topk}
    {w_1}A^2_1(t) - {w_i} A^2_i(t) \geq 1, \forall i \in \{k+1,...,N \}.
\end{equation}
This is because $w_i A^2_i(t) \in \mathbb{Z}^{+}$ for all sources $i$. Further, note that $f(x) = \sqrt{x}$ is Lipschitz for $x \in [1,\infty)$ with the Lipschitz constant $0.5$. Applying this fact to \eqref{eq:unique}, we get
\begin{equation}
    \label{eq:topk_sqrt}
    \sqrt{w_1}A_1(t) - \sqrt{w_i} A_i(t) \geq 0.5, \forall i \in \{k+1,...,N \}.
\end{equation}

Using these inequalities above, we obtain
\begin{equation}
\gamma_j \leq \frac{\sum_{i=1}^{N} \sqrt{w_i}}{\min_{j}\sqrt{w_j}} \big| \gamma_1 \big|, \forall j\in \{k+1,...,N \}. 
\end{equation}
Finally putting all of the inequalities together, we get 
\begin{align}
    \begin{aligned}
    \label{eq:exp_sum_topk}
    &\sum\limits_{j=1}^{N} \alpha^{w_j A^2_j(t)} \bigg( \sum_{i=1}^{N} \sqrt{w_i} \big(\sqrt{w_i} A_i(t) -  \sqrt{w_j} A_j(t)\big)   \bigg) \\
    &\leq k\alpha^{w_1 A^2_1(t)} \big( \gamma_1 \big)  + \sum_{i=k+1}^{N} \alpha^{w_{k+1} A^2_{k+1}(t)} \frac{\sum_{i=1}^{N} \sqrt{w_i}}{\min_{j}\sqrt{w_j}} \big| \gamma_1 \big| \\
    &\leq \alpha^{w_1 A^2_1(t)} \big| \gamma_1 \big| \bigg( -k + \alpha^{-1}\frac{(N-k)\sum_{i=1}^{N} \sqrt{w_i}}{\min_{j}\sqrt{w_j}} \bigg).  
    \end{aligned}
\end{align}
Again, choosing 
\begin{equation}
    \alpha > \bigg( \frac{(N-1)\sum_{i=1}^{N} \sqrt{w_i}}{\min_{j}\sqrt{w_j}} \bigg)
\end{equation}
is sufficient to guarantee that \eqref{eq:exp_sum_topk} is negative and hence \eqref{eq:drift_diff_1} is negative. This completes the proof of Lemma~\ref{lem:drift}.
\end{IEEEproof}

Next, we proceed to the proof of Theorem~\ref{thm:csma_opt}. The one-slot drift for the optimal stationary randomized policy is given by - \begin{align}
\label{eq:drift_sr_thm}
    \begin{aligned}
    \mathbb{E}&\bigg[\Delta^{sr}(t) \bigg| \bm{A(t)}\bigg] \\ = &\sum_{j=1}^{N} \pi^*_j \bigg( \sqrt{w_j} + \sum_{i \neq j} \sqrt{w_i} (A_i(t)+1)  \bigg) - \sum_{j=1}^{N} \sqrt{w_j} A_j(t)\\
    =& \sum_{j=1}^{N} \sqrt{w_j} - \sum_{j=1}^{N} \pi^{*}_j\sqrt{w_j}A_j(t).
    \end{aligned}
\end{align}

Putting together \eqref{eq:drift_comp} and \eqref{eq:drift_sr_thm}, we get
\begin{equation}
\label{eq:csma_drift_ub}
    \mathbb{E} \bigg[\Delta^{csma}(t) \bigg| \bm{A(t)}\bigg] \leq
    \sum_{j=1}^{N} \sqrt{w_j} - \sum_{j=1}^{N} \pi^{*}_j\sqrt{w_j}A_j(t).
\end{equation}
Summing \eqref{eq:csma_drift_ub} for $t = 1,...,T$ and taking expectation, we get
\begin{align}
    \begin{aligned}
    \mathbb{E}&\bigg[ L(T+1) -L(1) \bigg] \leq \\
     &T\sum_{j=1}^{N} \sqrt{w_j} - \sum_{j=1}^{N} \mathbb{E} \bigg[ \sum_{t=1}^{T}  \sqrt{w_j} \pi^{*}_j A_j(t) \bigg] .
    \end{aligned}
\end{align}
Substituting the expression for $\pi^{*}_j$ from \eqref{eq:opt_sr}, rearranging and dividing by $T$, we get
\begin{align}
    \begin{aligned}
    \frac{1}{T \big(\sum_{j=1}^{N} \sqrt{w_j}\big)}&\mathbb{E}\bigg[ \sum_{t=1}^{T} \sum_{j=1}^{N} w_j A_j(t) \bigg]   \leq \\ &\sum_{j=1}^{N} \sqrt{w_j} - \frac{1}{T} \mathbb{E}\bigg[ L(T+1) - L(1) \bigg]. 
    \end{aligned}
\end{align}
Since $L(T+1) \geq 0$ and $\frac{w_j}{\pi^*_j} = \sqrt{w_j} (\sum_{j=1}^{N} \sqrt{w_j})$, we can further simplify the equaiton above to
\begin{equation}
 \frac{1}{T}   \mathbb{E}\bigg[ \sum_{t=1}^{T} \sum_{j=1}^{N} w_j A_j(t) \bigg] \leq \sum_{j=1}^{N} \frac{w_j}{\pi^*_j} + \frac{\mathbb{E}[L(1)]}{T}.
\end{equation}
Now, we observe that the average AoI of a source $j$ under the optimal stationary randomized policy is given by $\bar{A}^{sr}_j = \frac{1}{\pi^{*}_j}$, as shown in \cite{kadota2018scheduling}. Using this fact and taking limsup as $T$ goes to infinity in the equation above, we get
\begin{equation}
    \sum_{j=1}^{N} w_j \bar{A}^{csma}_{j}(t) \leq \sum_{j=1}^{N} w_j \bar{A}^{sr}_j.
\end{equation}
We also know from \cite{kadota2018scheduling} that stationary randomized policies can be at most a factor of two away from optimal. Thus, we get
\begin{equation}
    \frac{ \sum_{i=1}^{N} w_i \bar{A}^{csma}_i }{ \sum_{i=1}^{N} w_i \bar{A}^{opt}_i } \leq  \frac{ \sum_{i=1}^{N} w_i \bar{A}^{sr}_i }{ \sum_{i=1}^{N} w_i \bar{A}^{opt}_i } \leq 2.
\end{equation}
This completes the proof.


\end{IEEEproof}

The factor of two optimality guarantee that we have derived is the same performance guarantee as the one shown for the max-weight policy in \cite{AoI_book} and better than the factor of four bound derived in \cite{kadota2018scheduling,talak2018distributed}. Viewing Theorems \ref{thm:csma_timeslot} and \ref{thm:csma_opt} together, we conclude that the idealized Fresh-CSMA policy can replicate the behavior of the max-weight policy, both at each time-slot with high probability and in terms of long-term average AoI over the entire time-horizon if $\alpha > \max\bigg((N-1)\frac{1-\delta}{\delta},  \frac{(N-1)\sum_{i=1}^{N} \sqrt{w_i}}{\min_{j}\sqrt{w_j}} \bigg)$. In Section \ref{sec:sim}, we will see via simulations that this holds true even for small values of $\alpha$, i.e. $\alpha$ doesn't need to be very large for Fresh-CSMA to be able to mimic the max-weight policy.




\section{Near-Realistic Multiple Access Model}
\label{sec:semi_real}
Until now, we have looked at distributed multiple access with idealized assumptions. In this section, we discuss the Fresh-CSMA protocol under a more realistic version of multiple access.

\begin{figure}
	\centering
	\includegraphics[width=0.99\linewidth]{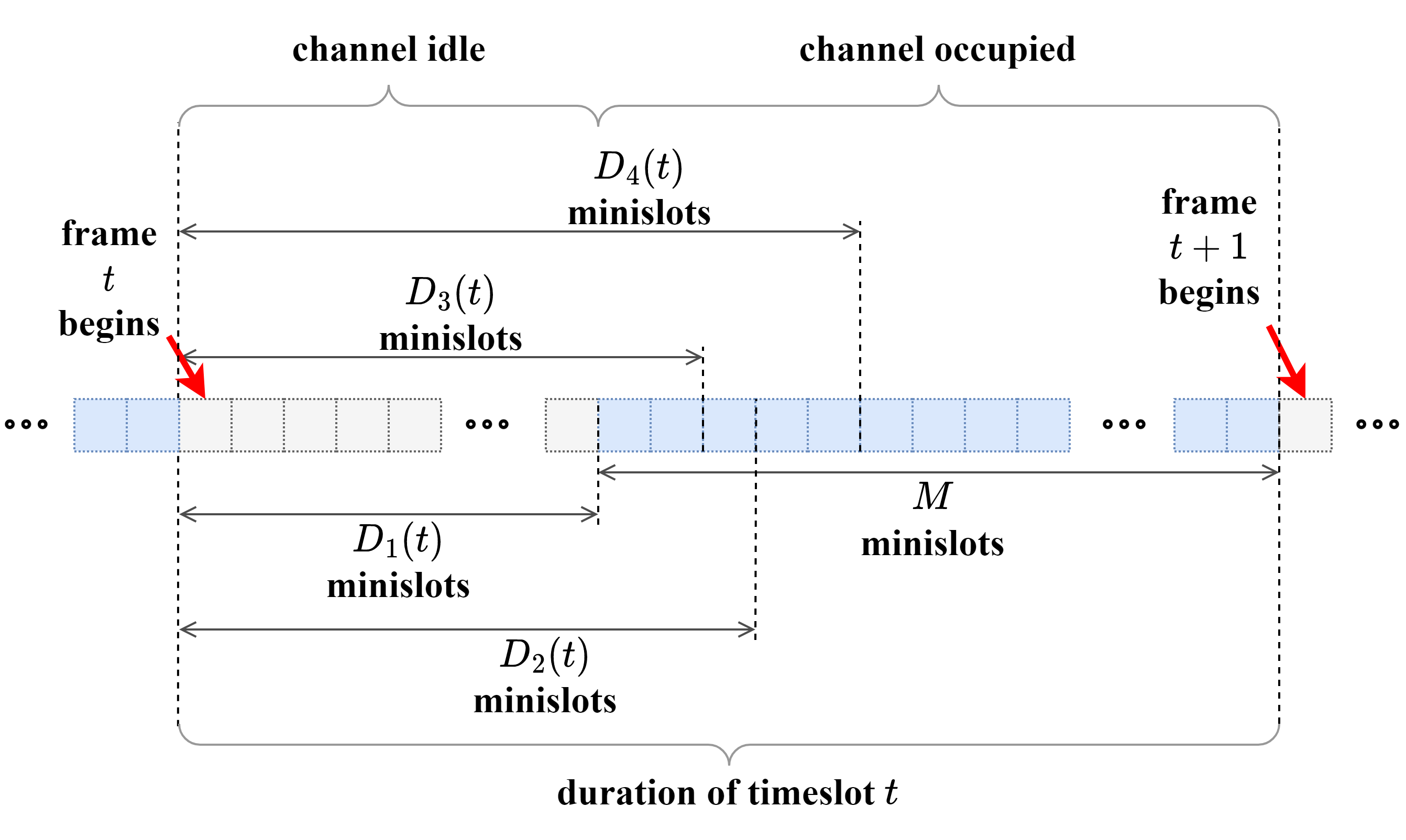}
	\caption{Events within frame $t$ in the near-realistic multiple access model. Four sources choose backoff timers $D_1(t),...,D_4(t)$. Source $1$'s timer runs out first, after which it transmits its update for $M$ minislots.}
	\label{fig:slots}
\end{figure}
Our discussion is based on the IEEE 802.11 standard for wireless LAN \cite{ieee802-11}. This standard defines a distributed coordination function (DCF) for sharing access to the wireless medium based on a CSMA/CA style protocol. The 802.11 standard divides time into the basic units of mini-slots, where each mini-slot is the duration of time needed by a source to detect packet transmission from any another source, i.e. perform channel sensing. A typical value for the mini-slot duration in IEEE 802.11 g/n/ac protocols is $9 \mu s$. 

Fig.~\ref{fig:slots} describes the key elements of our near-realistic model. First, we consider that the backoff timers for source $i$ at frame $t$, denoted by $D_i(t)$, can only run in multiples of minislot durations. This relaxes Assumptions \ref{ass:timers} and \ref{ass:sensing}, since the timers are now discrete, with finite precision and limited by the amount of time required to do channel sensing. Further, since timers are no longer continuous, the probability of collision is also non-zero and has an effect on the average age.

As before, we assume that backoff timers for each source begin at the beginning of each frame and the source/sources whose timers run out first get to transmit an entire application layer update. Thus, we have not relaxed Assumption~\ref{ass:slotted} in this model. We assume that a mini-slot takes $\frac{1}{M}$ units of time. So, transmitting an entire update takes $M$ mini-slots. We will discuss in Section \ref{sec:sim} that for typical values of update sizes and transmission rates $M$ tends to be large (around $10000$).

Finally, we also relax Assumption~\ref{ass:overhead} and consider the amount of time that the channel remains idle in each frame. Consider the example frame depicted in Fig.~\ref{fig:slots} with four backoff timers $D_1(t),...,D_4(t)$. The timer of source $1$, denoted by $D_1(t)$, runs out first. Then, source $1$ transmits its update for $M$ minislots. Thus the total duration of frame $t$ is $D_1(t) + M$ minislots or alternatively $1+\frac{D_1(t)}{M}$. However, for the first $D_1(t)$ minislots in frame $t$, the channel remained idle. We term this the \textit{backoff overhead}. In general, given $D(t) \triangleq \underset{i \in [N]}{\operatorname{min}}~~  D_i(t)$, frame $t$ takes $M + D(t)$ mini-slots or alternatively $1+\frac{D(t)}{M}$ units of time to complete. Compared to the idealized setting, the time $\frac{D(t)}{M}$ is the backoff overhead of the protocol, since it is the time that the channel must remain idle before any source starts transmitting. We take this overhead into account while calculating AoIs.

Our new model thus relaxes Assumptions \ref{ass:timers}, \ref{ass:sensing} and \ref{ass:overhead}, while allowing us to study the effect of collision probabilities and backoff overheads on the AoI. For this near-realistic multiple access model, we provide a modified version of the Fresh-CSMA protocol below. As before, We use $t$ to denote the discrete frames and $\tau$ to denote time within each frame, denoting the number of minislots that have passed within this frame.

\begin{algorithm}
	\DontPrintSemicolon
	\SetKwInOut{Input}{Input}\SetKwInOut{Output}{Output}
	\Input{parameters $\alpha > 1$, $\beta > 1, B \in \mathbb{Z}^{+}$}
	\While{ $t \in 1,...,T$ }{
		\For{$i \in 1,...,N$}{
		Generate a  random variable $Z_i(t) \sim \text{exp}\bigg(\alpha^{w_i A^2_i(t)}\bigg)$.\; 
		Map it to a non-negative integer timer,  $D_i(t) = \max \bigg(B + \big\lfloor \log_{\beta}\big(Z_i(t)\big) \big\rfloor,0\bigg)$.\;
		\While{$\tau < D_i(t)$}{
		Stay silent\;
		}
		\If{Channel is free}{
		Transmit\;
		}
		}
	}
	\caption{Near-Realistic Fresh-CSMA}
	\label{alg:s-Fresh-CSMA}
\end{algorithm}

The key difference between the protocol described in Alg.~\ref{alg:s-Fresh-CSMA} and Alg.~\ref{alg:i-Fresh-CSMA} is mapping the continuous random variables $Z_i(t)$ to the discrete variables $D_i(t) \in \{0\cup\mathbb{Z}^{+}\}$, which denote the number of mini-slots source $i$ should count down before transmitting. 
\subsection{Collisions}
Note that when using the protocol above, a packet collision happens if two sources $i$ and $j$ choose the same discrete backoff timers $D_i(t)$ and $D_j(t)$ at frame $t$ while also being the first timers to count down to zero, i.e. $\underset{j \in [N]}{\operatorname{argmin}}\bigg(  D_j(t) \bigg)$ is not unique. When a collision happens, we assume that the base station fails to receive an update from any of the transmitting sources and the entire frame is wasted. The following theorem analyzes the probability of the event that two sources $i$ and $j$ choose different backoff timers at time $t$.
\begin{framed}
\begin{theorem}
\label{thm:collision_prob}
Let the AoIs at time $t$ be given by $A_1(t),...,A_N(t)$ and $\lambda_i \triangleq \alpha^{w_i A^2_i(t)}$. Then probability that any two sources $i$ and $j$ choose different backoff timers $D_i(t)$ and $D_j(t)$ can be lower bounded as follows
\begin{align}
\begin{aligned}
    \mathbb{P}\bigg(D_i(t) \neq D_j(t)\bigg) \geq \psi\big(B,\beta, \lambda_i, \lambda_j\big) + \psi\big(B,\beta, \lambda_j, \lambda_i\big)
\end{aligned}
\end{align}
where, the function $\psi(\cdot)$ is given by:
\begin{align*}
    \psi\big(B,\beta, \lambda_i, \lambda_j\big) = \frac{\lambda_i e^{-\beta^{-B}(\lambda_i + \beta \lambda_j)}}{\lambda_i + \beta \lambda_j} \\ + \bigg(e^{\lambda_i \beta^{-B}} - 1\bigg)e^{-\beta^{-B}(\lambda_i + \beta \lambda_j)}.
\end{align*}
\end{theorem}
\end{framed}
\begin{IEEEproof}
Consider the exponential random variables generated by the two sources $Z_i(t)\sim\exp(\lambda_i)$ and $Z_j(t)\sim\exp(\lambda_j)$, where $\lambda_i = \alpha^{w_i A^2_i(t)}, \forall i$. Suppose that the following inequality holds:
\begin{equation}
\label{eq:two_source_c1}
    Z_j(t) > \begin{cases}
         \beta Z_i(t), &\text{ if } Z_i(t) > \beta^{-B}\\
         \beta^{-B+1}, &\text{ otherwise.}
    \end{cases}
\end{equation}
Then it is easy to see that \[
    \max\big(B + \big\lfloor \log_{\beta}\big(Z_i(t)\big) \big\rfloor,0\big) < \max\big(B + \big\lfloor \log_{\beta}\big(Z_j(t)\big) \big\rfloor,0\big)
\] which in turn implies that $D_i(t) < D_j(t)$. Switching $i$ and $j$ in the inequality \eqref{eq:two_source_c1}, we get $D_i(t) > D_j(t)$. Note that the two events are disjoint.

Thus, we can lower-bound our probability of interest as follows:
\begin{align}
    \begin{aligned}
    \label{eq:th3_eq1}
    \mathbb{P}(&D_i(t) \neq D_j(t))  
    \geq\mathbb{P}\bigg( Z_j(t) > \min\{\beta Z_i(t), \beta^{-B+1}\}\bigg) \\ &+ \mathbb{P}\bigg(Z_i(t) > \min\{\beta Z_j(t), \beta^{-B+1}\} \bigg)
\end{aligned}
\end{align}

Simplifying the first term on the RHS, we get
\begin{align}
    \begin{aligned}
    \mathbb{P}&\bigg( Z_j(t) > \min\{\beta Z_i(t), \beta^{-B+1}\}\bigg) \\
    =& \int_{0}^{\beta^{-B}} \lambda_i e^{-\lambda_i x} e^{-\lambda_j \beta^{-B+1}} dx + \int_{\beta^{-B}}^{\infty} \lambda_i e^{-\lambda_i x} e^{-\lambda_j \beta x} dx\\
    =& \frac{\lambda_i e^{-\beta^{-B}(\lambda_i + \beta \lambda_j)}}{\lambda_i + \beta \lambda_j} + \bigg(e^{\lambda_i \beta^{-B}} - 1\bigg)e^{-\beta^{-B}(\lambda_i + \beta \lambda_j)}\\
     \triangleq &\psi\big(B,\beta, \lambda_i, \lambda_j\big).
    \end{aligned}
\end{align}
By the same argument, the second term on the RHS of \eqref{eq:th3_eq1} is equal to $\psi\big(B,\beta, \lambda_j, \lambda_i\big)$. Thus, we get
\begin{equation}
    \mathbb{P}(D_i(t) \neq D_j(t)) \geq \psi\big(B,\beta, \lambda_i, \lambda_j\big) + \psi\big(B,\beta, \lambda_j, \lambda_i\big).
\end{equation}
This completes the proof.
\end{IEEEproof}

As a corollary of this proof, note that
\begin{align}
\begin{aligned}
    \frac{\partial}{\partial B}   &\psi\big(B,\beta, \lambda_i, \lambda_j\big) \\
    &= \lambda_j \beta^{-B+1} \log(\beta) \bigg(e^{\lambda_i \beta^{-B}} - 1\bigg)e^{-\beta^{-B}(\lambda_i + \beta \lambda_j)}\\
    &\geq 0.
\end{aligned}
\end{align}
The last inequality follows since $\beta > 1$ and $\lambda_i > 0$. Thus, the probability that two sources choose different backoff timers \textit{increases with the parameter $B$}. In the limit as $B\rightarrow\infty$, we get
\begin{equation}
\label{eq:inf_B}
    \lim_{B\rightarrow\infty} \mathbb{P}(D_i(t) \neq D_j(t)) \geq  \frac{\lambda_i}{\lambda_i + \beta\lambda_j} + \frac{\lambda_j}{\lambda_j + \beta\lambda_i}.
\end{equation}
Thus, when $B$ is large, the probability that two sources occupy different mini-slots \textit{decreases with $\beta$}. Putting the two observations together, we should choose a large value for $B$ and a small value for $\beta$ (close to $1$) to reduce collisions. However, for finite $B$, \eqref{eq:inf_B} does not hold and the value of $\beta$ cannot be too small, since in that case, all the discrete timers will map to the first minislot leading to collisions in almost every frame. 

In the analysis above, we used the probability of two sources occupying different backoff timers as a proxy for analyzing the collision probability directly, since a tight bound for the actual collision probability is too involved to compute. In Section \ref{sec:sim}, we will see via simulations how the actual collision probability varies with the parameters $B$ and $\beta$.

\subsection{Backoff Timer Overhead}
Next, we analyze the overhead of the backoff timers in the Fresh-CSMA protocol in the near-realistic model. This is unlike the idealized setting where we ignored the time taken by the backoff timers to count down, during which the channel remains idle. 

Recall that the quantity $\frac{D(t)}{M}$ is what we defined as the backoff overhead of a protocol at time $t$, since it is the time that the channel must remain idle before any source starts transmitting. The following theorem provides an upper-bound on the expected backoff overhead in frame $t$. 
\begin{framed}
\begin{theorem}
\label{thm:timer_overhead}
Let the AoIs at time $t$ be $A_1(t),...,A_N(t)$ and $\lambda_i \triangleq \alpha^{w_i A^2_i(t)}$. Then, the expected idle-time of the Fresh-CSMA protocol at time $t$ can be upper-bounded by
\begin{equation}
    \frac{1}{M}\mathbb{E}\big[D(t)\big] \leq \frac{1}{M} + \frac{\Gamma\big(0,\lambda \beta^{-B}\big)}{M\log(\beta)}.
\end{equation}
Here $\Gamma(\cdot,\cdot)$ is the upper incomplete gamma function and $\lambda \triangleq \sum_{i \in [N]} \lambda_i$.
\end{theorem}
\end{framed}
\begin{IEEEproof}
Let $Z(t) = \min_{i\in[N]}Z_i(t)$. Since $Z(t)$ is the minimum of $N$ independent exponential random variables, it is also exponentially distributed with the parameter $\lambda = \sum_{i\in[N]}\lambda_i$. Using this, we provide an upper bound for $\mathbb{E}[D(t)]$ below.
\begin{align}
    \begin{aligned}
    \mathbb{E}\big[D(t)\big] &= \mathbb{E}\bigg[ \min_{i \in [N]} \big(D_i(t)\big)   \bigg]
    \\
    &= \mathbb{E}\bigg[ \min_{i \in [N]} \bigg( \max \big(B + \big\lfloor \log_{\beta}\big(Z_i(t)\big) \big\rfloor,0\big) \bigg)  \bigg]\\
    &\leq B+1+\mathbb{E}\bigg[ \max\bigg(\log_{\beta}\big( \min_{i\in[N]} Z_i(t) \big),-B\bigg) \bigg]\\
    &\leq B+1+\mathbb{E}\bigg[ \max\bigg(\log_{\beta}\big( Z(t) \big),-B\bigg) \bigg]\\
    &\leq 1 + \frac{\Gamma\big(0,\lambda \beta^{-B}\big)}{\log(\beta)}.
    \end{aligned}
\end{align}
The last inequality follows from the fact that $\mathbb{E}\big[\max(\log(Z(t)),-B)\big] = -B + \Gamma\big(0,\lambda \beta^{-B}\big)$, where $\Gamma(\cdot,\cdot)$ is the upper incomplete gamma function given by
\begin{equation}
    \Gamma(s,x) = \int_{x}^{\infty} t^{s-1} e^{-t} dt.
\end{equation}
\end{IEEEproof}
Note that the function $\Gamma(0,x)$ is given by
\begin{equation}
\label{eq:Gamma}
    \Gamma(0,x) = \int_{x}^{\infty} t^{-1} e^{-t} dt.
\end{equation}
From \eqref{eq:Gamma}, note that $\Gamma(0,x)$ is decreasing in $x$. Thus, for a fixed value of $\beta$, the expected idle time \textit{increases as $B$ increases}. The exact dependence on $\beta$ is more tricky to evaluate, so we again consider the case of large $B$ as an alternative. First, we make the following observation for the gamma function of large values of $B$
\begin{equation}
    \Gamma(0,\lambda \beta^{-B}) \approx B \log(\beta) - log(\lambda) - \gamma,
\end{equation}
where $\gamma \approx 0.58$ is the Euler-Mascheroni constant. Using this, it is easy to see that for large values of $B$, the expected idle time upper bound is approximately equal to $1 + B - \log_{\beta}(\lambda)$. Thus, the backoff overhead also \textit{increases with $\beta$}, given a fixed large value of $B$. 

Together this implies that we need to choose a relatively small value of $B$ and a small value of $\beta$ (close to $1$) to reduce idle time. Importantly, there is a tradeoff between the collision probability and the backoff overhead depending on the choice of the parameter $B$. A larger value of $B$ reduces the probability of collision but at the cost of higher backoff overhead. 

Theorem~\ref{thm:timer_overhead} also allows us to compute an approximate upper-bound for the average idle time over the entire horizon. Suppose the average AoI of source $i$ under the Fresh-CSMA policy in the near-realistic model is denoted by $\bar{A}_i$. Using the average AoIs, we define the following quantity: $ \bar{\lambda} \triangleq \sum_{i=1}^{N} \alpha^{w_i \bar{A}^2_i}. $
Then, an approximate upper-bound of the average backoff overhead per frame over the entire time-horizon $\bar{D}^{ub}$ can be obtained as follows:
\begin{equation}
\label{eq:approx_overhead}
    \bar{D}^{ub} \approx \frac{1}{M} + \frac{\Gamma\big(0,\bar{\lambda} \beta^{-B}\big)}{M\log(\beta)}.
\end{equation}
In Section~\ref{sec:sim}, we will see via simulations that this is a good bound for the average backoff overhead in the near-realistic setting.

\section{Going Beyond Age of Information}
\label{sec:AoII}
While Age of Information has been used as a proxy for optimizing monitoring and control costs in real-time settings over the past decade, a recent line of work starting with \cite{maatouk2020age} has proposed a new and more general metric to measure the impact of stale information on underlying real-time monitoring tasks. This metric is called the Age of Incorrect Information (AoII), and it has been used to study the monitoring of Markov sources in various kinds of settings \cite{maatouk2020age, kam2020age, kriouile2021minimizing, maatouk2022age}. 

The key idea behind the AoII metric is that takes into account the actual error or distortion between the estimate of the process being monitored at the remote monitor and the actual value of the process at present. More precisely, suppose that $X(t)$ represents the state of the process that needs to be monitored and let $\hat{X}(t)$ be the estimate of the process at time $t$ at the remote monitor. Let the function $g(X(t),\hat{X}(t))$ measure the distortion or error between the actual process and its remote estimate and let $f(\cdot)$ be a monotone increasing function. Let $V(t)$ represent the most recent time instant up to the current time $t$ at which this distortion was zero. The AoII is then defined as
\begin{equation}
	\label{eq:AoII-def}
	AoII(t) \triangleq f\big(t-V(t)\big) g(X(t),\hat{X}(t)).
\end{equation}

Note that if the actual process and its estimate stay the same for some time despite no new updates being delivered to the monitor, the AoII remains zero while the AoI keeps increasing. Thus, the AoII can be viewed as a more accurate metric to measure information uncertainty at the monitor.

Now, consider a setting with $N$ sources, sending updates to the base station, where only one source can talk to the base station at any given time. Unlike the setting we have analyzed until now, we will now look at minimizing the sum of AoIIs instead of weighted AoIs. Each source is tracking a process $X_i(t), i \in \{1,...,N\}$ and the base station maintains estimates for each process $\hat{X}_i(t), i \in \{1,...,N\}$. Using these estimates and \ref{eq:AoII-def}, we can compute the AoIIs for each source, denoted by $AoII_i(t), i \in \{1,...,N\}$. We want to design a scheduling policy that minimizes the long term time-average of the AoIIs, i.e.
\begin{equation}
	\label{eq:AoII_opt_prob}
	\underset{\pi}{\operatorname{argmin}} \bigg( \limsup_{T \rightarrow \infty}   \bigg[\frac{1}{T} \frac{1}{N} \sum_{t = 1}^{T} \sum_{i=1}^{N} AoII_{i}(t) \bigg] \bigg).
\end{equation}

A good candidate policy to solve this problem would be to schedule the source with the highest AoII at each time-slot. 
\begin{equation}
	\label{eq:max_AoII}
	\pi^{max-AoII}(t) = \underset{j \in [N]}{\operatorname{argmax}}\bigg(  AoII_j(t)\bigg).
\end{equation}

However, a crucial drawback of the AoII metric is the fact that its computation requires knowledge of the actual current state of the process $X(t)$. Thus, AoIIs cannot be computed at the base station beforehand and a centralized multiple source scheduling policy like \eqref{eq:max_AoII} cannot be implemented in reality, since the base station does not know the actual current states of each process. The CSMA based protocols we develop in this work provide a way out of this dilemma. Sources can compute their own AoIIs, since they have access to $X(t)$, $\hat{X}(t)$, $t$ and $V(t)$. Then, a CSMA style policy that uses AoIIs instead of the AoIs can be implemented to pick the source that has the highest AoII and get better monitoring performance. We illustrate how to incorporate AoII into a CSMA style policy using the idealized version of Fresh-CSMA in Alg.~\ref{alg:i-Fresh-CSMA-AoII}. The near-realistic version follows immediately, by replacing the weighted AoI with the AoII.
\begin{algorithm}
	\DontPrintSemicolon
	\SetKwInOut{Input}{Input}\SetKwInOut{Output}{Output}
	\Input{parameters $\alpha > 1, \delta \ll 1$}
	\While{ $t \in 1,...,T$ }{
		\For{$i \in 1,...,N$}{
			Generate a  random timer $Z_i(t) \sim \text{exp}\bigg(\alpha^{AoII_i(t)}\bigg)$.\; 
			\While{$\tau < \delta Z_i(t)$}{
				Stay silent\;
			}
			\If{Channel is free}{
				Transmit\;
			}
		}
	}
	\caption{Idealized Fresh-CSMA with AoIIs}
	\label{alg:i-Fresh-CSMA-AoII}
\end{algorithm}

For the setting involving monitoring Markov sources, the following choice of the functions $f(\cdot)$ and $g(\cdot)$ is typically considered in literature
\begin{equation}
	\label{eq:AoII_markov}
	AoII(t) = \big(t-V(t)\big) \mathbbm{1}_{ \{ X(t) \neq \hat{X}(t)\} }.
\end{equation}
For this specific AoII metric, we can show a result similar to Theorem~\ref{thm:csma_timeslot} in the case of weighted AoI.
\begin{framed}
	\begin{theorem}
		\label{thm:csma_timeslot_AoII}
		Given any $\delta \in (0,1)$ and $AoII_1(t),...,AoII_N(t)$ evolving according to \ref{eq:AoII_markov}; if we set $\alpha \geq (N-1)\frac{1-\delta}{\delta}$, then the following holds
		\begin{equation}
			\mathbb{P}\bigg(\pi^{CSMA-AoII}(t) = \pi^{max-AoII}(t)\bigg) \geq 1-\delta.
		\end{equation}
		Here, $\pi^{max-AoII}(t)$ is the scheduling decision given by \eqref{eq:max_AoII}, while $\pi^{CSMA-AoII}(t)$ is the scheduling decision made by the idealized Fresh-CSMA policy that utilizes AoIIs (Alg.~\ref{alg:i-Fresh-CSMA-AoII})
	\end{theorem}
\end{framed}
\begin{IEEEproof}
	The proof is identical to that of Theorem~\ref{thm:csma_timeslot} - all we need to do is replace the weighted AoIs with the AoIIs. The evolution of AoII according to \ref{eq:AoII_markov} ensures that all the AoII values are integers. As before, we divide the proof into two parts.
	
	\textbf{Case 1:} The expression $\underset{j \in [N]}{\operatorname{argmax}}\bigg(  AoII_j(t)\bigg)$ has a unique maximum. Let this maximum be the source $1$ without loss of generality. Then, the max-AoII decision is to schedule source $1$. Since we know that AoIIs are integers by \eqref{eq:AoII_markov}, so the following must hold:
	\begin{equation}
		AoII_1(t) - AoII_i(t) \geq 1, \forall i \neq 1.
		\label{eq:th5_c1_ineq}
	\end{equation}

	Now, applying Lemma 1, we can calculate the following probability
	\begin{align}
		\begin{aligned}
			\mathbb{P}\bigg(\pi^{CSMA-AoII}(t) = 1\bigg) &= \frac{\alpha^{AoII_1(t)}}{\sum\limits_{j=1}^{N}\alpha^{AoII_j(t)} }\\
			&= \frac{1}{ 1 + \sum_{i=2}^{N} \alpha^{AoII_i(t) - AoII_1(t)}}\\
			&\geq \frac{1}{1+(N-1)\alpha^{-1}}\\
			&\geq 1-\delta.
		\end{aligned}
	\end{align}
	The first inequality follows by using \eqref{eq:th5_c1_ineq} while the second inequality follows due to the fact that $\alpha \geq (N-1)\frac{1-\delta}{\delta}$.
	
	\textbf{Case 2:} The expression $\underset{j \in [N]}{\operatorname{argmax}}\bigg(  AoII_j(t)\bigg)$ has multiple maxima. 
	Suppose that the set of maxima is given by the nodes $\{1,...,k\}$. Then, the max-AoII policy will choose one of these sources to be scheduled. We want to lower bound the probability that Fresh-CSMA with AoIIs chooses a node from within this set. To do so, we first make a similar observation as in the case above.
	\begin{equation}
		AoII_1(t) - AoII_j(t) \geq 1,  \forall j = k+1,...,N.
		\label{eq:th5_c2_ineq}
	\end{equation}
	Using Lemma 1, we calculate the probability of interest
	\begin{align}
		\begin{aligned}
			\mathbb{P}\bigg( \pi^{CSMA-AoII}(t) &\in \{1,...,k \} \bigg) = \frac{k \alpha^{AoII_1(t)}}{\sum\limits_{j=1}^{N}\alpha^{AoII_j(t)} } \\
			&= \frac{k}{ k + \sum_{i=k+1}^{N} \alpha^{AoII_i(t) - AoII_1(t)}}\\
			&\geq \frac{1}{1+(N-k)\alpha^{-1}}\\
			&\geq 1-\delta.
		\end{aligned}
	\end{align}
	As before, the first inequality follows by using \eqref{eq:th5_c2_ineq} while the second inequality follows due to the fact that $\alpha \geq (N-1)\frac{1-\delta}{\delta}$. This completes the proof.
\end{IEEEproof}

Theorem~\ref{thm:csma_timeslot_AoII} shows that on a per-time-slot basis, Fresh-CSMA based on AoII matches the scheduling decisions made by the hypothetical centralized max-AoII policy \eqref{eq:max_AoII}, which cannot be implemented in reality. In Section~\ref{sec:sim}, we will show via simulations that the Fresh-CSMA policy based on AoII can achieve lower time-average AoII across the network than even the centralized max-weight policy that utilizes only AoIs. This suggests that in certain settings distributed policies that utilize monitoring error can outperform centralized policies that have to make scheduling decisions while being oblivious to the actual processes. 

\section{Numerical Results}
\label{sec:sim}
\begin{figure}
	\centering
	\includegraphics[width=0.99\linewidth]{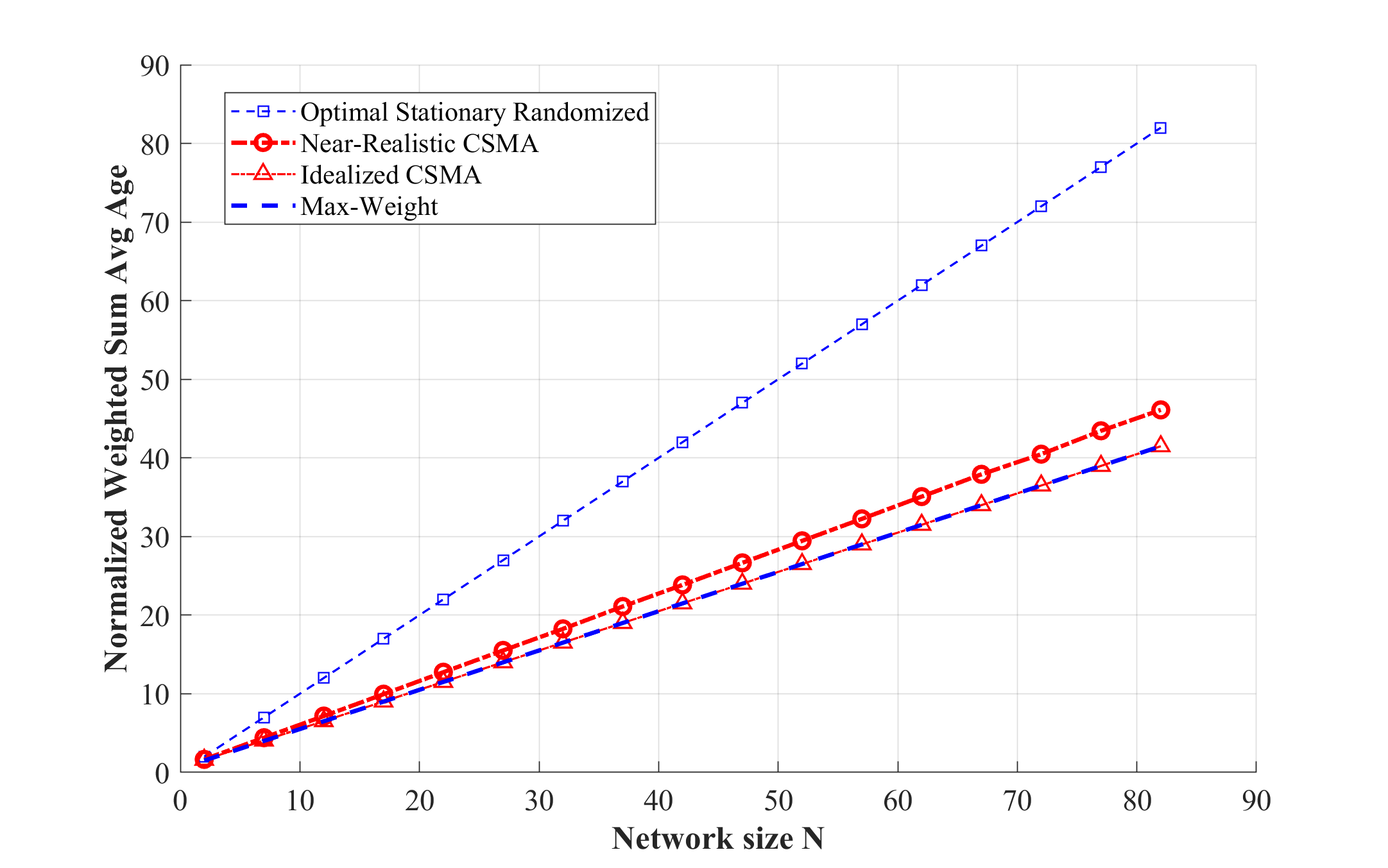}
	\caption{Normalized average AoI vs system size ($N$) for symmetric weights.}
	\label{fig:equal_weights}
\end{figure}

\begin{figure}
	\centering
	\includegraphics[width=0.99\linewidth]{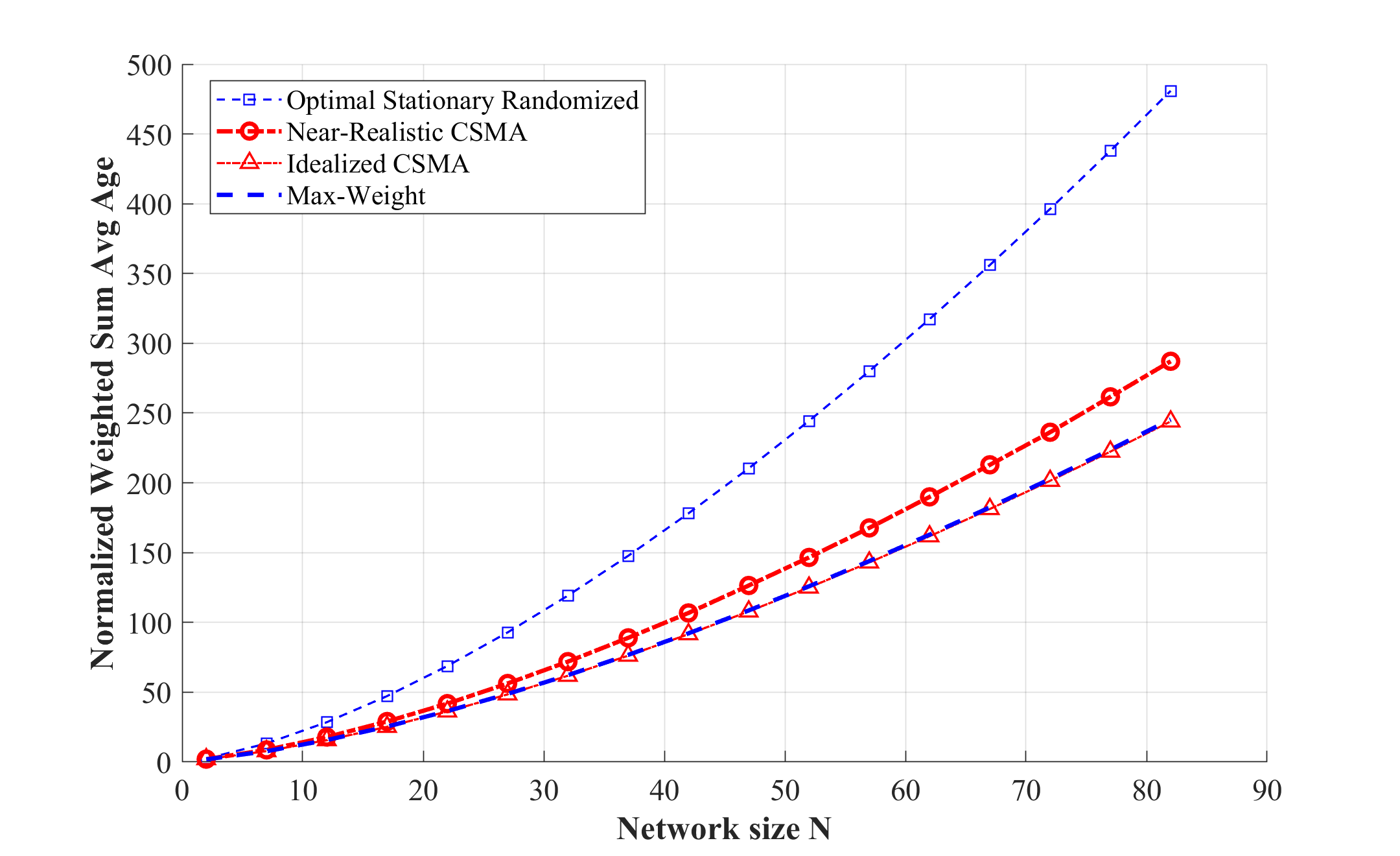}
	\caption{Normalized average AoI vs system size ($N$) for unequal weights.}
	\label{fig:unequal_weights}
\end{figure}
To verify the theoretical results developed in earlier sections, we now provide numerical results from packet level simulations of all the policies. Throughout this section, we assume that the minislot is $9\mu s$ long (a typical value in IEEE 802.11 implementations \cite{ieee802-11}) and an update packet from a each source is roughly 600 kB.

Thus, at a data rate of 54 Mbps (which is the highest data rate for IEEE 802.11g in the 2.4 GHz band \cite{ieee802-11}), it takes 10000 minislots to finish sending an update. This, in turn, implies that we set $M=10000$ for our simulations, i.e. each packet takes 10000 minislots to transmit.

Each experiment involves calculating the time-average of AoI, AoII, collision probabilities or backoff timer overheads. All of these time-averages are reported for experiments that involve 100000 application layer update packets being transmitted. Unless otherwise specified, we set $\alpha = 1 + \frac{1}{\sum_i {w_i}}$, $\beta = 1.1 + \max(\log( \log (N)) , 0)$, and $B = 250 + N$ for the near-realistic CSMA implementation.

First, we consider the weighted sum average age minimization problem in the setting with equal weights, i.e. $w_i = 1, \forall i$. Fig.~\ref{fig:equal_weights} plots the performance of two centralized policies - the optimal stationary randomized policy and the max-weight policy, as well as two distributed policies - the idealized Fresh-CSMA protocol and the near-realistic Fresh-CSMA protocol, as the the number of sources in the system $N$ increases. We plot the normalized weighted sum average age, i.e. $\frac{1}{N} \sum_{i} w_i \bar{A}_i$ for each policy. We observe that the idealized Fresh-CSMA protocol matches the performance of the max-weight policy almost exactly, as expected from Theorems \ref{thm:csma_timeslot} and \ref{thm:csma_opt}. Further, the near-realistic Fresh-CSMA protocol has only a small performance gap to the max-weight policy (due to collisions and backoff overheads) but still significantly outperforms the optimal stationary randomized policy. 

Next, we consider the same setting but with asymmetric weights. We set the weight for source $k$ to be $\sqrt{k}$, i.e. $w_k = \sqrt{k}$, and plot the normalized average age as the system size $N$ increases in Fig.~\ref{fig:unequal_weights}. We make the same observations regarding the performance of the policies as in the case of symmetric weights and further note that while our theoretical results needed weights $w_i$ to be integers, that assumption is not required to get good performance in practice.

\begin{figure}
	\centering
	\includegraphics[width=0.99\linewidth]{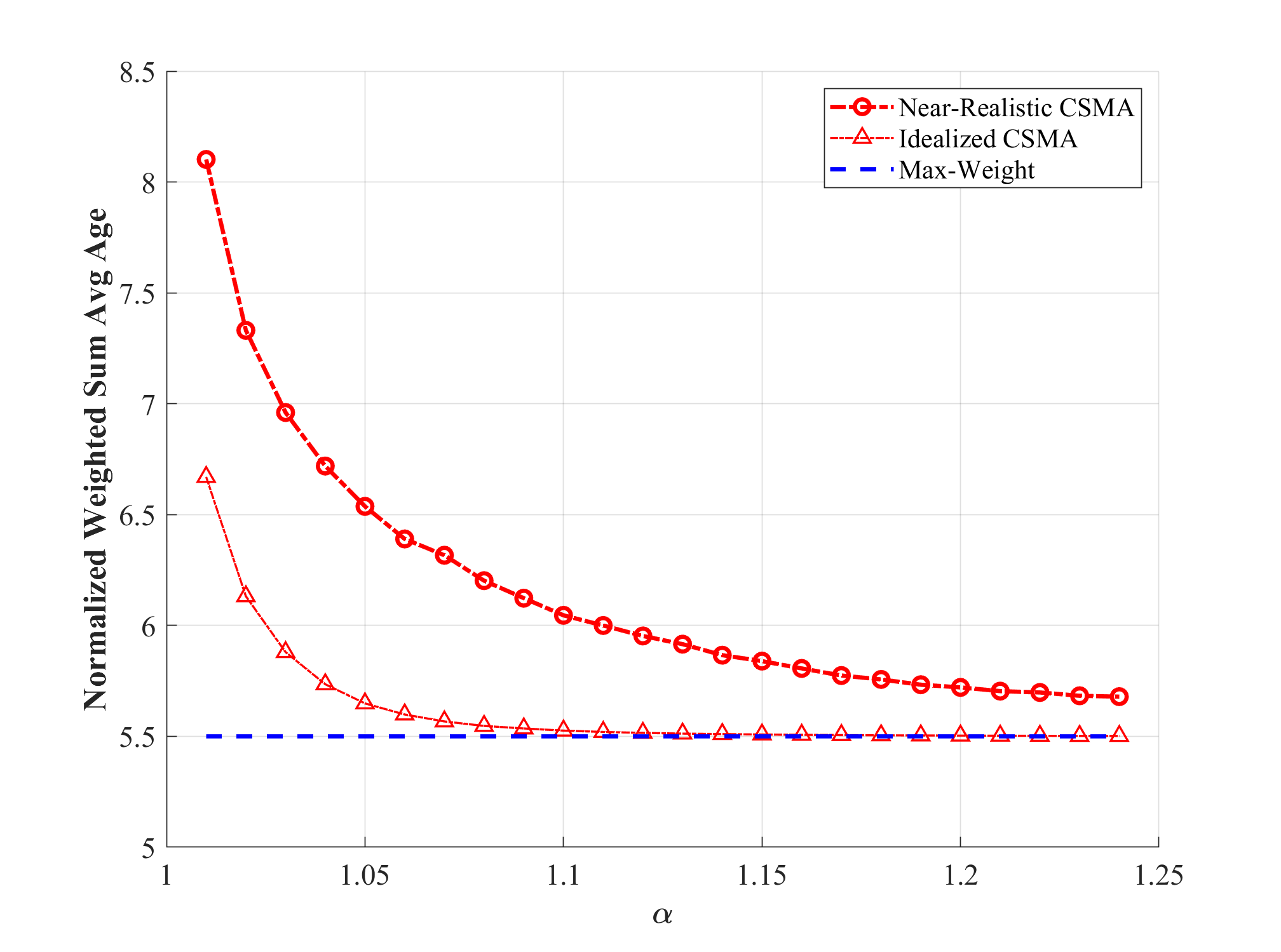}
	\caption{Normalized average AoI vs $\alpha$.}
	\label{fig:AoI_alpha}
\end{figure}
Next, we consider how the performance of our proposed protocols depends on the parameter $\alpha$. Theorems \ref{thm:csma_timeslot} and \ref{thm:csma_opt} suggest that $\alpha$ needs to be very large to guarantee performance of Fresh-CSMA close to max-weight. However, we find that Fresh-CSMA starts performing very similarly to max-weight even at very small values of $\alpha$, as show in Fig.~\ref{fig:AoI_alpha} for a symmetric system with $N=10$ sources. This is important in practice since large values of $\alpha$ could lead to integer overflows. As expected, the performance gap narrows as $\alpha$ increases.
  
\begin{figure}
	\centering
	\includegraphics[width=0.95\linewidth]{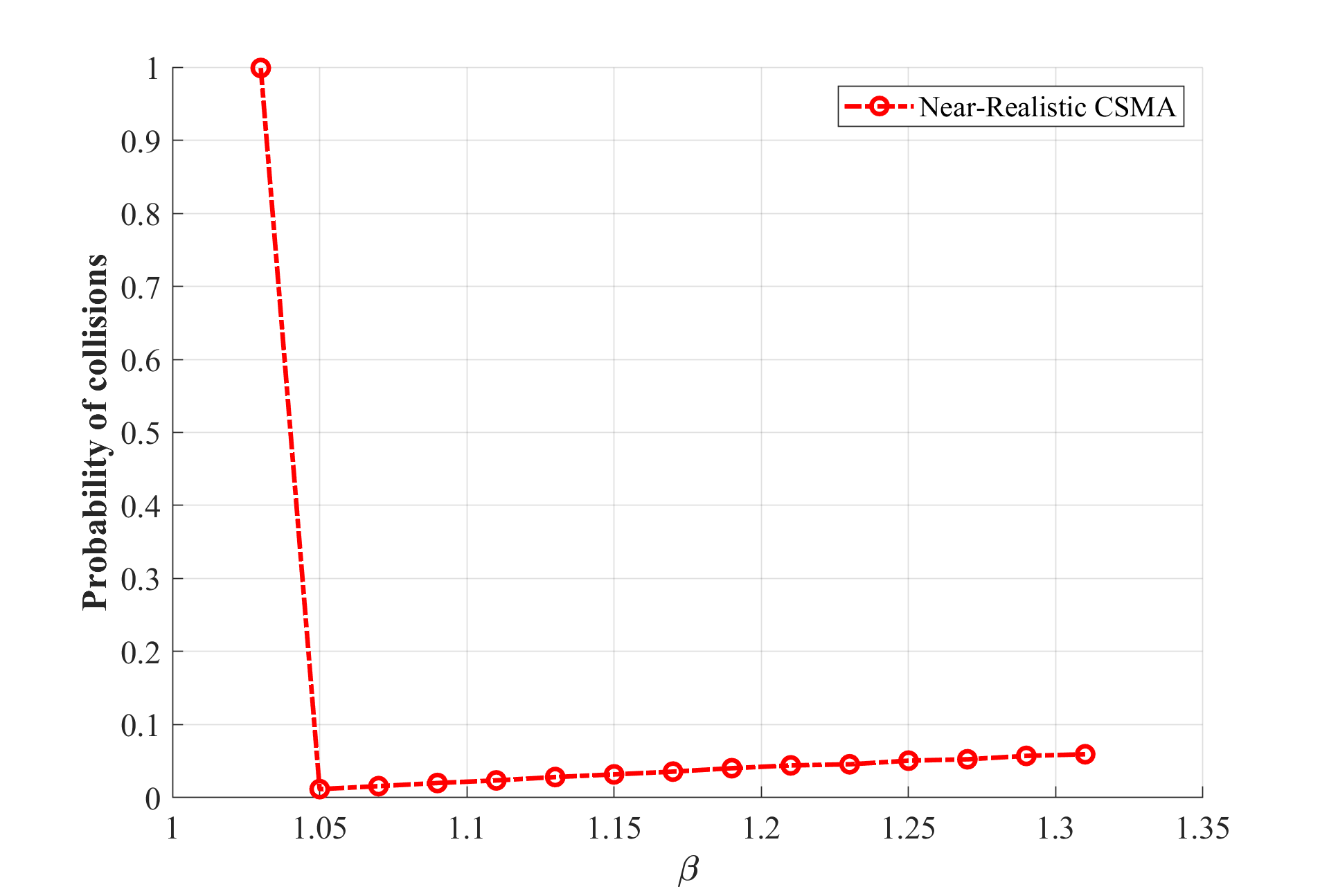}
	\caption{Probability of collisions vs $\beta$}
	\label{fig:coll_beta}
\end{figure}

\begin{figure}
	\centering
	\includegraphics[width=0.99\linewidth]{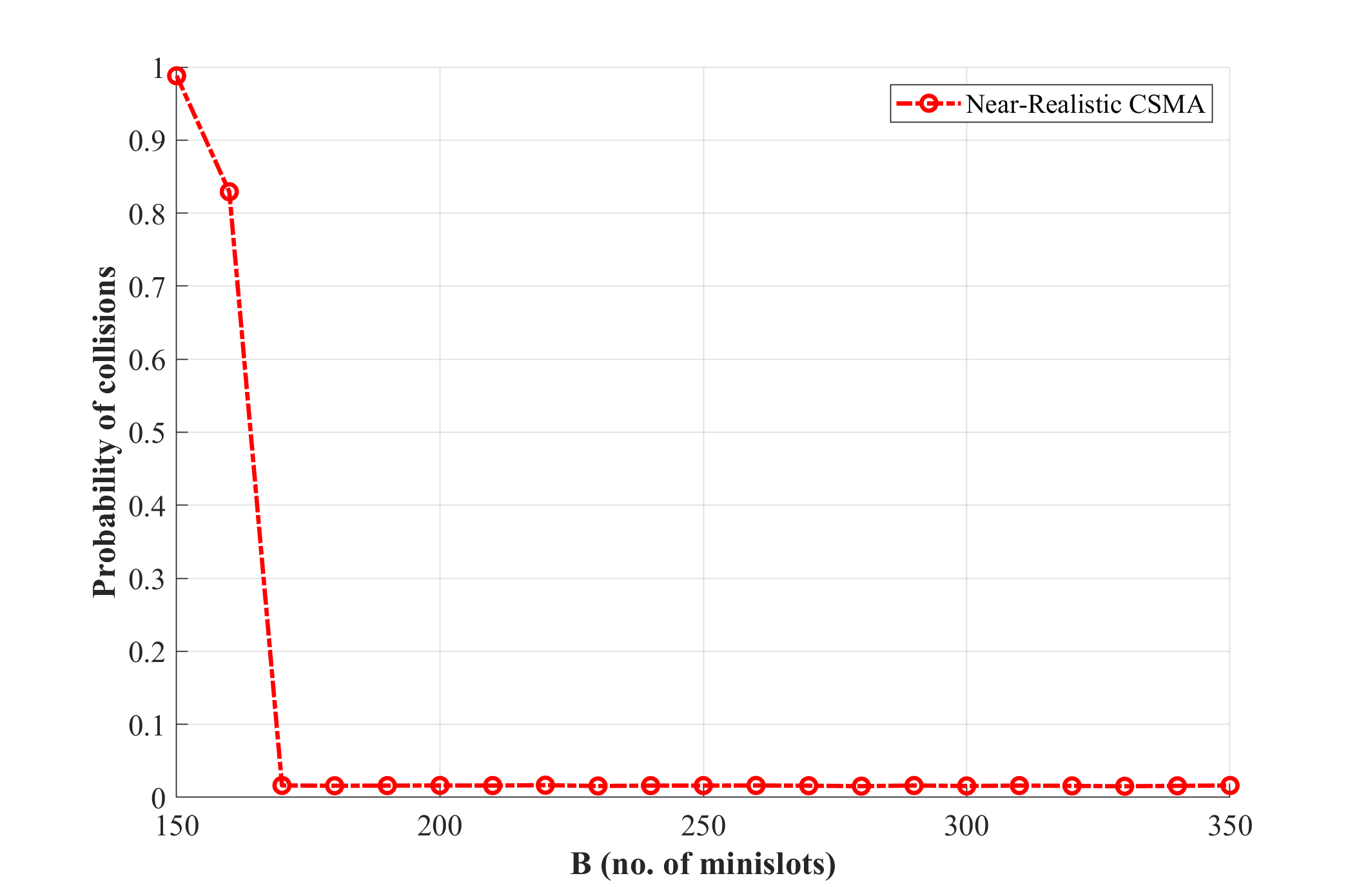}
	\caption{Probability of collisions vs $B$}
	\label{fig:coll_B}
\end{figure}

Next, we look at the near-realistic Fresh-CSMA protocol in detail. In Fig.~\ref{fig:coll_beta}, we plot the collision probability for this protocol in a symmetric system with $N=10$ sources as $\beta$ is varied,  while fixing all other parameters. We observe that for $\beta < 1.05$, the collision probability is $1$ since all timers map to the first mini-slot. However, as we increase $\beta$ beyond $1.05$, we observe that the collision probability first drops to $0.01$ and then gradually increases to $0.06$. In Fig.~\ref{fig:coll_B}, we plot the collision probability as $B$ is varied, while fixing all other parameters. For small values of $B$, the collision probability is almost $1$, but as $B$ increases beyond a threshold, the collision probability stays roughly constant at around $0.015$. These results are in line with what we expected from Theorem \ref{thm:collision_prob}.

\begin{figure}
	\centering
	\includegraphics[width=0.99\linewidth]{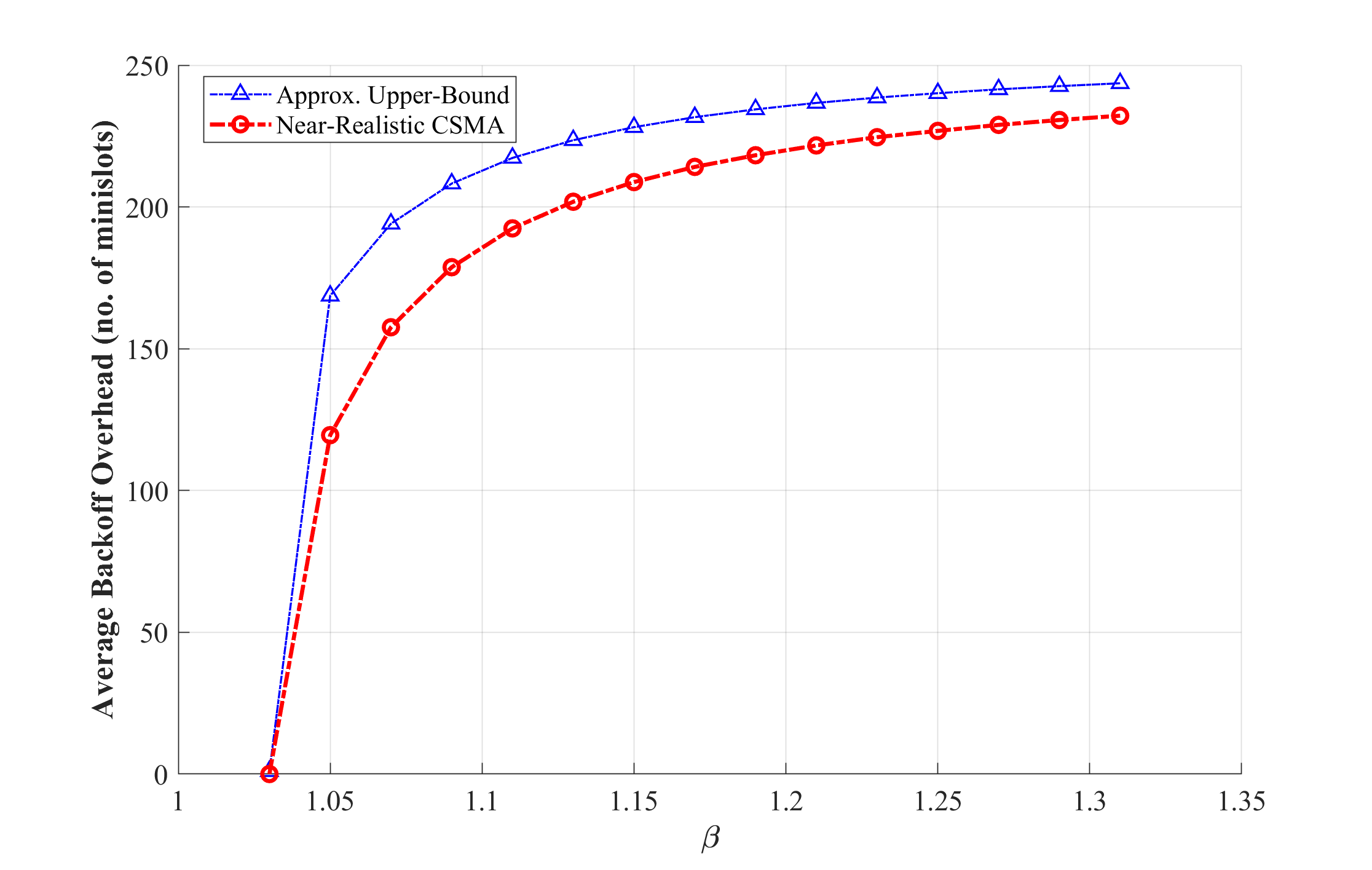}
	\caption{Average backoff overhead  vs $\beta$}
	\label{fig:overhead_beta}
\end{figure}
\begin{figure}
	\centering
	\includegraphics[width=0.99\linewidth]{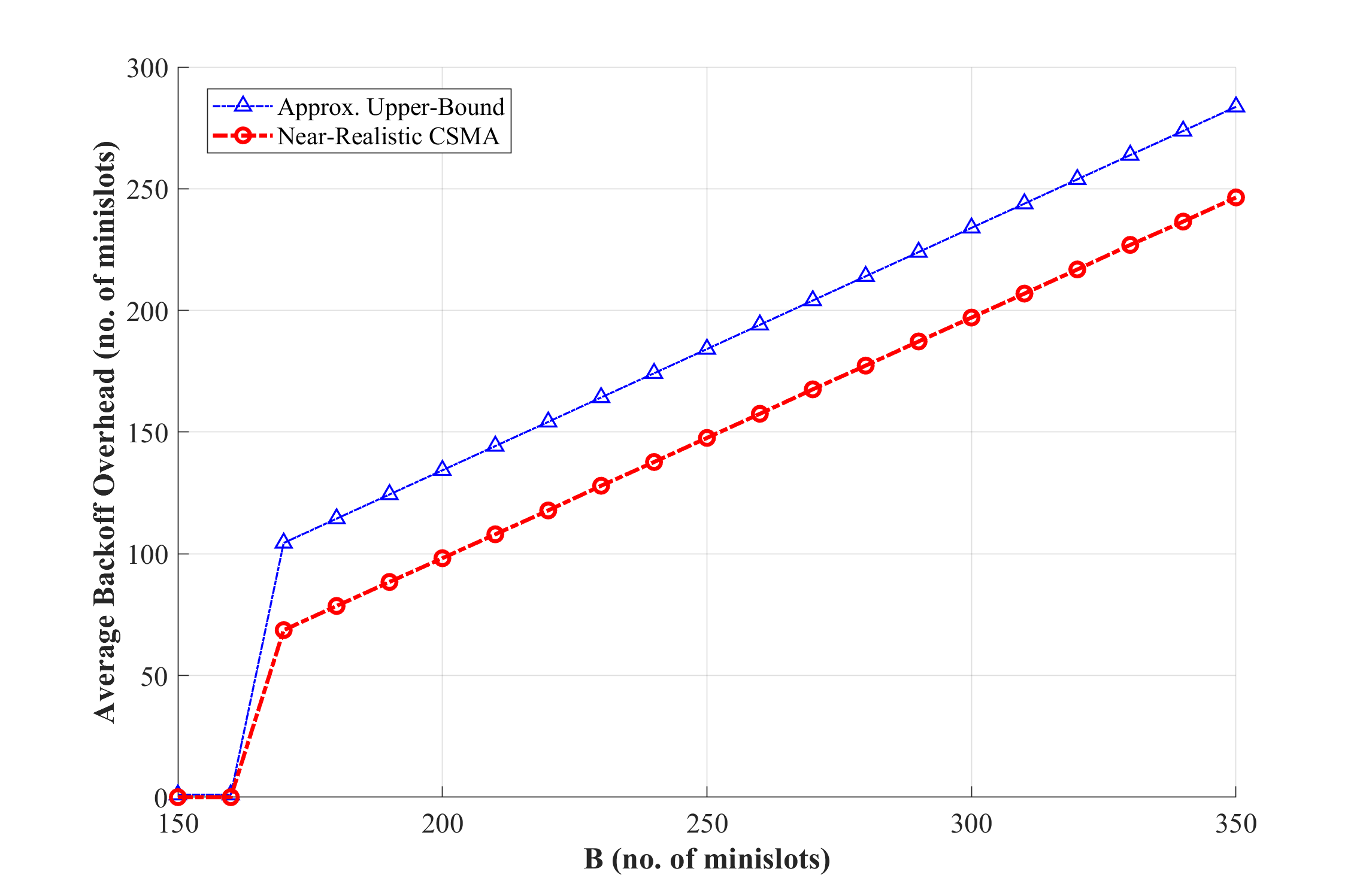}
	\caption{Average backoff overhead  vs $B$}
	\label{fig:overhead_B}
\end{figure}
\begin{figure}
	\centering
	\includegraphics[width=0.5\linewidth]{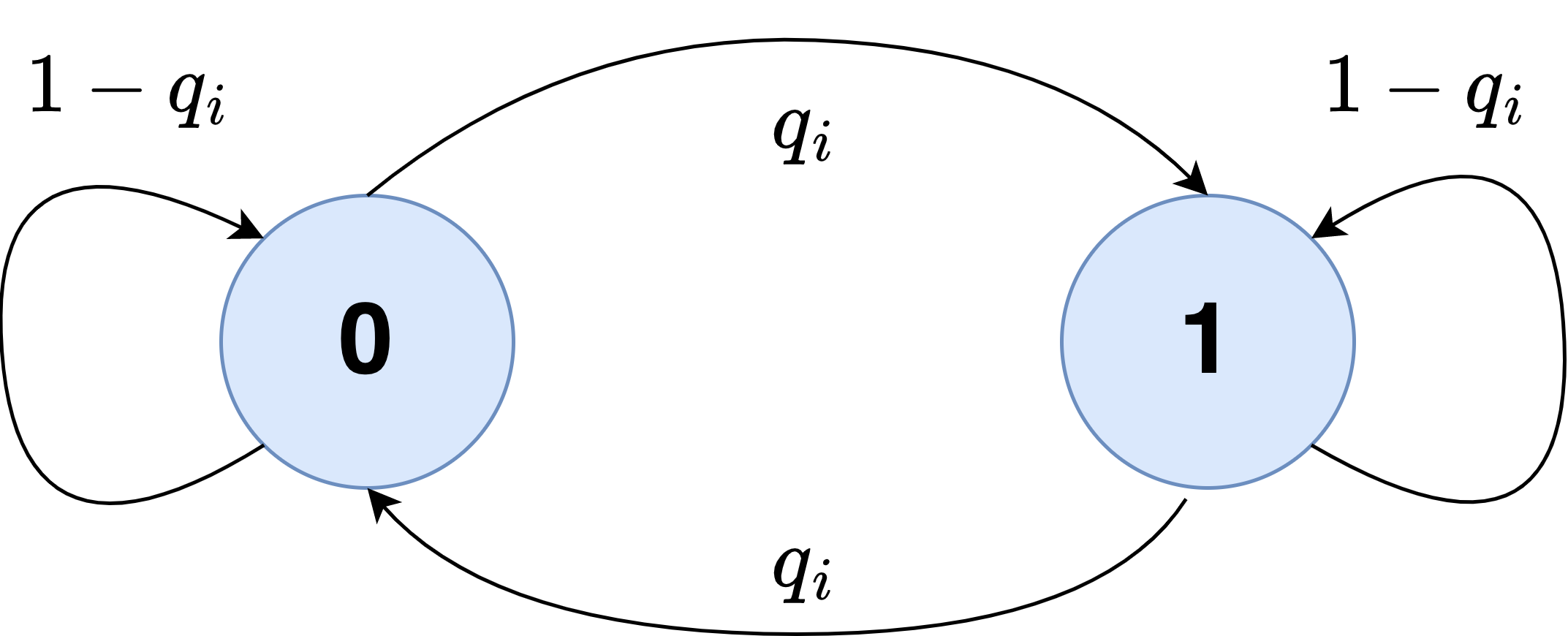}
	\caption{Symmetric two-state Markov chain representing the $i$th source.}
	\label{fig:markov_chain}
\end{figure}
\begin{figure}
	\centering
	\includegraphics[width=\linewidth]{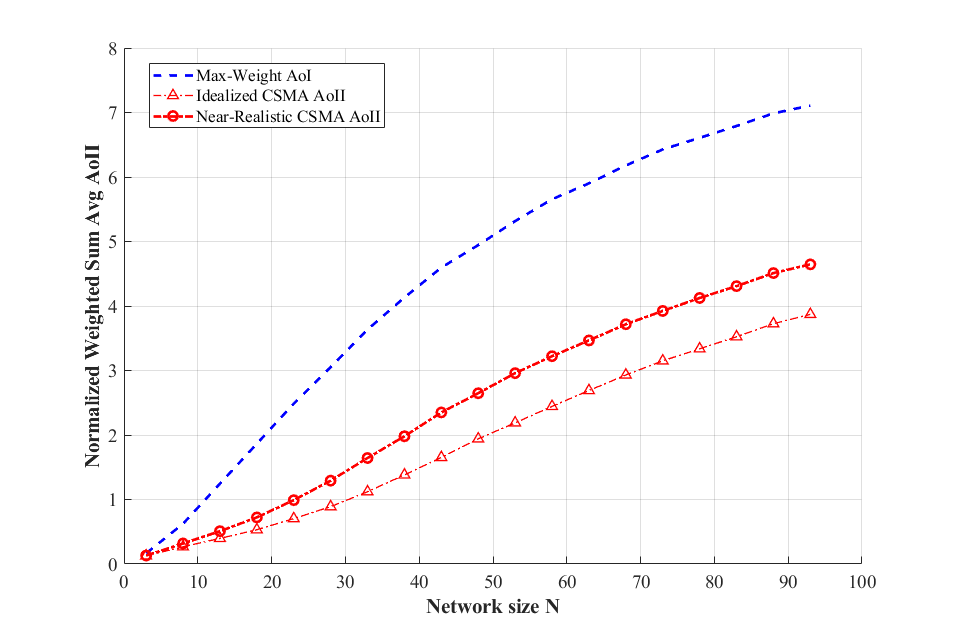}
	\caption{Normalized average AoII vs system size $N$ when symmetric Markov sources}
	\label{fig:AoII_N}
\end{figure}
\begin{figure}
	\centering
	\includegraphics[width=0.99\linewidth]{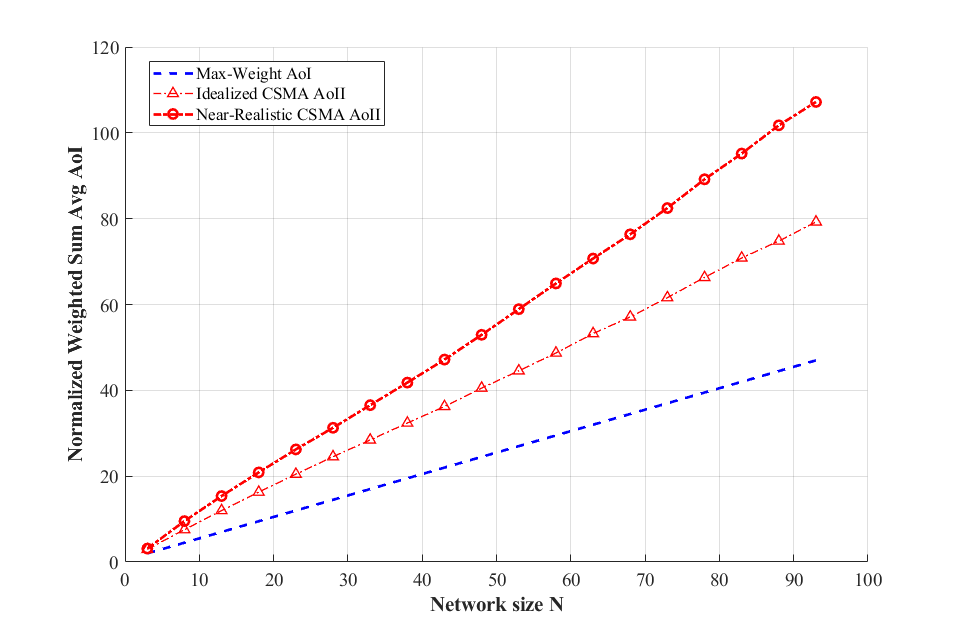}
	\caption{Normalized average AoI vs system size $N$ when monitoring symmetric Markov sources}
	\label{fig:AoII_N_AoI}
\end{figure}
Next, we look at the average overhead of near-realistic Fresh-CSMA. Fig.~\ref{fig:overhead_beta} and Fig.~\ref{fig:overhead_B} plot the average overhead (in number of minislots) as the parameters $\beta$ and $B$ are varied, respectively. As expected from Theorem \ref{thm:timer_overhead}, the overhead increases with both $\beta$ and $B$. We note that our approximate expression for the overhead \eqref{eq:approx_overhead} is a good upper-bound that can be used in practice for system design. Also, we observe that the overhead remains relatively small (2-3\%) compared to the update size, which takes $10000$ minislots.

Finally, we consider the AoII metric discussed in Section~\ref{sec:AoII}. In this context, we look at the setting where each source is a symmetric two-state Markov chain, evolving independently over time (see Fig.~\ref{fig:markov_chain}). We set the transition probability $q_i$ for each source to be $0.05$ and implement the following three policies - centralized max-weight that uses AoI, distributed idealized Fresh-CSMA that uses AoII and distributed near-realistic Fresh-CSMA that uses AoII. For the CSMA implementations, we set the parameters as follows $\alpha = 2.1, \beta = 1.05 + \log(\log (N))$ and $B = 250 + \lfloor N/4 \rfloor$.

Fig.~\ref{fig:AoII_N} plots the time-average AoII performance of the three policies as we increase the number of sources in the system $N$. Here AoII is computed using \eqref{eq:AoII_markov}. Clearly, the distributed CSMA versions deliver much better AoII performance than the centralized policy that is only able to utilize AoIs. Specifically, for $93$ source, the idealized Fresh-CSMA with AoII performs about $45\%$ better than AoI max-weight and the near-realistic version of Fresh-CSMA with AoII performs about $35\%$ better than AoI max-weight. This suggests that our CSMA based design is general and can easily accommodate other kinds of information freshness and distortion metrics.

Interestingly, the gain in AoII performance comes at the cost of higher AoIs. In Fig.~\ref{fig:AoII_N_AoI}, we plot the long-term time-average AoIs for the same three policies while monitoring Markov sources as the number of source $N$ is increased. We observe that the distributed CSMA versions have higher AoIs than the centralized AoI max-weight. So, Fresh-CSMA based on AoII trades off better monitoring performance/error measured in terms of AoII with worse performance in terms of standard AoI.

\section{Conclusion}
In this work, we designed a distributed CSMA protocol to minimize weighted sum Age of Information in single-hop wireless networks. We showed that under idealized assumptions, our proposed protocol can closely replicate the behavior of centralized policies known to be nearly optimal. We also analyzed our protocol under a near-realistic medium access model and showed how system parameter choices affect packet collisions and overhead. Our simulation results confirm that our protocol works well in practice and that the performance gap between the idealized version and the near-realistic version of Fresh-CSMA is small. We have also extended some of our results to AoII, a more general information freshness metric. Two important directions of work involve further analysis of our protocols beyond AoI and AoII to metrics such as monitoring error or real-time control costs, and implementing the protocol in real systems to compare performance against standard WiFi.

\bibliographystyle{ieeetr}
\bibliography{bibliography_2}

\end{document}